\pdfoutput=1 
\documentclass[a4paper, 11pt]{article}
\PassOptionsToPackage{table}{xcolor}

\usepackage{jheppub} 

\usepackage[T1]{fontenc} 
\usepackage{tangocolors}
\usepackage{tikz}
\usepackage{tikz-3dplot}
\usetikzlibrary{arrows, decorations,backgrounds, patterns}
\usetikzlibrary{decorations.pathreplacing ,decorations.markings}
\usepackage{amssymb}
\usetikzlibrary{decorations.pathmorphing,backgrounds,shapes,arrows,shadows}
\usetikzlibrary{patterns.meta}
\usetikzlibrary{patterns,decorations.pathmorphing}
\tikzset{
    snake it/.style={decorate, decoration=snake}
}
\usepackage{pgfplots}
\pgfplotsset{compat=1.11}
\usepgfplotslibrary{fillbetween}
\usetikzlibrary{intersections}
\pgfdeclarelayer{bg}
\pgfsetlayers{bg,main}
\tikzset{zigzag/.style={decorate,decoration=zigzag}}
\tikzset{snake it/.style={decorate, decoration=snake}}
\makeatletter
\def\@hex@@Hex#1%
 {\if a#1A\else \if b#1B\else \if c#1C\else \if d#1D\else
  \if e#1E\else \if f#1F\else #1\fi\fi\fi\fi\fi\fi \@hex@Hex}
\makeatother

\usepackage[all]{xy}
\usepackage[percent]{overpic}
\usepackage{slashed}
\usepackage{wrapfig}
\usepackage{tabu}
\usepackage{diagbox}
\usepackage{mathrsfs,amsmath,amssymb,amsthm,amsfonts,tikz,graphicx,accents,hyperref, color}
\usepackage{dsfont,epiolmec, latexsym, stmaryrd, comment}
\usepackage{slashed,ccaption}
\usepackage{mathrsfs, calligra}
\usepackage{leftidx}
\usepackage{import}
\usepackage{multirow}
\usepackage{amsfonts}
\usepackage{pifont}
\usepackage{tabularx}
\usepackage{cancel}
\usepackage[utf8]{inputenc}
\usetikzlibrary{intersections,calc}
\usepackage{ifthen}
\usepackage{amsmath}
\usepackage{cancel}
\usepackage{caption} 
\usepackage{subcaption}

\usetikzlibrary{patterns.meta}
\usepackage{array}
\hypersetup{ linktoc=all,
    colorlinks, linkcolor={palatinateblue},
    citecolor={brightpink}, urlcolor={amaranth}
}

\graphicspath{{Images/}}

\renewcommand{\d}[1]{\ensuremath{\operatorname{d}\!{#1}}}

\def\sideremark#1{\ifvmode\leavevmode\fi\vadjust{\vbox to0pt{\vss
 \hbox to 0pt{\hskip\hsize\hskip1em
 \vbox{\hsize2cm\tiny\raggedright\pretolerance10000
 \noindent #1\hfill}\hss}\vbox to8pt{\vfil}\vss}}}%
\DeclareSymbolFont{extraup}{U}{zavm}{m}{n}
\DeclareMathSymbol{\varheart}{\mathalpha}{extraup}{86}
\DeclareMathSymbol{\vardiamond}{\mathalpha}{extraup}{87}
\makeatletter
\renewcommand*{\@fnsymbol}[1]{\ensuremath{\ifcase#1\or \clubsuit \or \vardiamond \or \varheart\or
    \spadesuit\or \mathparagraph\or \|\or **\or \dagger\dagger
    \or \ddagger\ddagger \else\@ctrerr\fi}}
\makeatother

\definecolor{rosy}{RGB}{230,235,252}
\definecolor{myframetitle}{RGB}{90,89,170}
\definecolor{myblocktitle}{RGB}{140,185,249}
\definecolor{mytitle}{RGB}{10,80,26}

\definecolor{darkgreen}{RGB}{27,130,45}
\definecolor{darkblue}{rgb}{0,0,0.3}
\definecolor{darkred}{rgb}{0.7,0,0}

\definecolor{light gray}{RGB}{220,220,220}
\definecolor{dark purple}{RGB}{108,0,217}
\definecolor{pink}{RGB}{190,20,100}
\definecolor{orang}{RGB}{193,63,0}
\definecolor{green}{RGB}{11,98,17}
\definecolor{darkpink}{RGB}{153,0,76}
\definecolor{bluegreen}{RGB}{0,102,102}
\definecolor{greenlagan}{RGB}{0,102,0}
\definecolor{redgreen}{RGB}{102,102,0}
\definecolor{Redgreen}{RGB}{153,76,0}
\definecolor{vividviolet}{rgb}{0.62, 0.0, 1.0}
\definecolor{amaranth}{rgb}{0.9, 0.17, 0.31}
\definecolor{palatinateblue}{rgb}{0.15, 0.23, 0.89}
\definecolor{brightpink}{rgb}{1.0, 0.0, 0.5}
\definecolor{cornflowerblue}{rgb}{0.39, 0.58, 0.93}
\definecolor{deepcarminepink}{rgb}{0.94, 0.19, 0.22}
\definecolor{radicalred}{rgb}{1.0, 0.21, 0.37}

\newcommand{\bTh}{\boldsymbol{\Theta }}

\newcommand{\bO}{\boldsymbol{\Omega}}

\DeclareFontFamily{OT1}{rsfs}{}

\DeclareFontShape{OT1}{rsfs}{m}{n}{ <-7> rsfs5 <7-10> rsfs7 <10->rsfs10}{} 

\DeclareMathAlphabet{\mycal}{OT1}{rsfs}{m}{n}

\newcommand{\be}{\begin{equation}}
\newcommand{\ee}{\end{equation}}
\newcommand{\bea}{\begin{eqnarray}}
\newcommand{\eea}{\end{eqnarray}}
\makeatletter \@addtoreset{equation}{section}

\begin{document}

\newcommand{\mytitle}{\begin{center}{\LARGE{\textbf{Heisenberg Soft Hair on Robinson-Trautman Spacetimes}}}
\end{center}}

\title{{\mytitle}}

\author[a,b]{H.~Adami}
\author[c]{, A. Parvizi}
\author[c]{, M.M.~Sheikh-Jabbari}
\author[c,d]{, V.~Taghiloo}

\affiliation{$^a$ Yau Mathematical Sciences Center, Tsinghua University, Beijing 100084, China}
\affiliation{$^b$ Beijing Institute of Mathematical Sciences and Applications (BIMSA), \\ Huairou District, Beijing 101408, P. R. China}
\affiliation{$^c$ School of Physics, Institute for Research in Fundamental
Sciences (IPM),\\ P.O.Box 19395-5531, Tehran, Iran}
\affiliation{$^d$ Department of Physics, Institute for Advanced Studies in Basic Sciences (IASBS),\\ 
P.O. Box 45137-66731, Zanjan, Iran}
\emailAdd{
hamed.adami@bimsa.cn, a.parvizi@ipm.ir, 
jabbari@theory.ipm.ac.ir, v.taghiloo@iasbs.ac.ir
}
\abstract{We study 4 dimensional $(4d$) gravitational waves (GWs) with compact wavefronts, generalizing Robinson-Trautman (RT) solutions in Einstein gravity with an arbitrary cosmological constant. We construct the most general solution of the GWs in the presence of a causal, timelike, or null boundary when the usual tensor modes are turned off. Our solution space besides the shape and topology of the wavefront which is a generic compact, smooth, and orientable $2d$ surface $\Sigma$, is specified by a vector over $\Sigma$ satisfying the conformal Killing equation and two scalars that are arbitrary functions over the causal boundary, the boundary modes (soft hair). We work out the symplectic form over the solution space using covariant phase space formalism and analyze the boundary symmetries and charges. The algebra of surface charges is a Heisenberg algebra. Only the overall size of the compact wavefront and not the details of its shape appears in the boundary symplectic form and is canonical conjugate to the overall mass of the GW.  Hence, the information about the shape of the wavefront can't be probed by the boundary observer.  We construct a boundary energy-momentum tensor and a boundary current, whose conservation yields the RT equation for both asymptotically AdS and flat spacetimes. The latter provides a hydrodynamic description for our RT solutions.
}
\maketitle
\section{Introduction}\label{sec:Intro}

Gravitational waves (GWs) are the latest observationally confirmed prediction of Einstein's gravity \cite{LIGOScientific:2016aoc}. Pure Einstein gravity (in the absence of matter fields) admits GW solutions. In 4 dimensional ($4d$) spacetimes, which will be the focus of this work, GWs have a 2 dimensional ($2d$) wavefronts. One can recognize two general classes of GWs,  planar wavefront GWs  (\textit{pp}-waves) \cite{kundtpp-wave} and those with compact wavefronts. The famous solutions of the former are  Brinkmann waves \cite{Brinkmann:1925fr} and the latter are the Robinson-Trautman (RT) \cite{Robinson:1960zzb, Robinson:1962zz} solutions. See a recent paper by Roger Penrose \cite{Penrose:2024fdd} for some historical remarks on these solutions. In this work we explore further the class of RT solutions.

The RT solutions are solutions to pure Einstein gravity admitting a shear and twist-free congruence of null geodesics with non-zero expansion \cite{Stephani:2003tm}.\footnote{In contrast,  \textit{pp}-waves admit shear and twist-free null geodesic congruence with zero expansion \cite{kundt1962study}.} The Goldberg-Sachs theorem \cite{Goldberg-Sachs} states that any vacuum Einstein solution with shear-free null congruence is algebraically special and falls into Petrov type II, III, N, D or O classes.  Einstein field equations for the RT family reduce to a single parabolic fourth-order differential equation, the RT equation. RT equation, as pointed out by K. Tod \cite{Tod_1989}, is the Calabi flow equation \cite{calabi1982extremal, calabi2002space}. The RT solutions have been constructed in de Sitter (dS) or anti-de Sitter (AdS), as well as flat backgrounds. From a physical perspective, the RT metrics can be interpreted as representing a system that is isolated and emits gravitational radiation which typically settles down to a Schwarzschild-like black hole. In this regard, this class of solutions has a specific property that allows one to study black holes and gravitational waves simultaneously. 

The causal and global structure, presence of past or future horizons, singularity, smoothness and analyticity of RT solutions have been discussed in the literature \cite{chrusciel1992global, Bicak:1997ne, Griffiths:2002gm,chrusciel1992non, Bicak:1995vc, Krtous:2003tc}. Notably, these studies reveal that RT wavefronts can exhibit diverse topologies, including spherical, toroidal, and higher genus configurations. Moreover, it appears that the RT Schwarzschild-like horizon (which in general is not a Killing horizon) is not analytic in time. Specifically, extending these spacetimes beyond the event horizon requires a finite degree of continuity. Moreover, it has been shown that for significantly general initial data on a codimension one null surface, and for transverse metrics diffeomorphic to a sphere, the RT solutions tend to converge to the static Schwarzschild solution after radiating for a significant duration \cite{lukacs1984lyapunov, Chrusciel:1991vxx}.

The study of the class of algebraically special solutions of Einstein's gravity with a negative cosmological constant, Petrov-type D classification of RT solutions,  reveals a profound correspondence between $2+1$ dimensional CFT living on the boundary and their bulk geometries \cite{BernardideFreitas:2014eoi, Gath:2015nxa, Mukhopadhyay:2013gja, Ciambelli:2017wou}. These solutions allow for the study of conformal, relativistic fluids and the extraction of higher-order transport coefficients. Furthermore, the characterization of perfect Cotton geometries elucidates the thermodynamic description of holographic theories, linking black hole uniqueness to holographic perfect equilibrium. Through the reconstruction of exact Einstein spacetimes, the relationship between boundary conditions and bulk geometries is manifest.

In a parallel line of developments, starting by the seminal work of Bondi et al and Sachs \cite{Bondi:1962px, Sachs:1962}, asymptotic symmetries of spacetimes which may involve GWs have been explored. The variation in the BMS charges (in asymptotically flat spacetimes) \cite{Barnich:2009se, Strominger:2017zoo} has been linked to the gravitational memory effects \cite{Strominger:2014pwa, Pasterski:2015tva}. In these analysis, certain falloff behavior for the metric of spacetime (and hence for GWs) are considered and moreover, the topology of the spatial part of the null infinity is assumed to be a two-sphere. The class of RT solutions do not necessarily respect either of these two assumptions. In addition, we may have RT solutions in asymptotic (A)dS backgrounds. 
Extension of asymptotic symmetry analysis  beyond standard BMS ones has been also considered, see e.g. 
\cite{Compere:2020lrt, Geiller:2022vto} and \cite{Fernandez-Alvarez:2020hsv, Fiorucci:2020xto, Poole:2021avh, Perez:2022jpr, Bonga:2023eml, Geiller:2024amx, Mao:2018xcw, Mao:2019ahc}. The boundary symmetry analysis is not limited to asymptotic regions of spacetime and has been extended to any boundary, any codimension 1 surface. In particular, boundary symmetries for null or timelike boundary have been analyzed \cite{Adami:2020amw, Adami:2021nnf, Adami:2021kvx, Adami:2023wbe, Sheikh-Jabbari:2022mqi, Chandrasekaran:2018aop, Chandrasekaran:2020wwn, Ciambelli:2023mir, Donnelly:2020xgu}. 

In this paper, our focus is on exploring the solution space of $4d$ RT metrics in presence of a generic causal boundary. We construct the solution space and analyze its inherent symmetries, the symplectic structure, and the associated surface charges. This investigation is a continuation of our previous studies, where we analyzed symmetries and charges near a generic null surface for $2d$ or $3d$ gravity theories, as discussed in \cite{Adami:2020ugu}. This analysis was later extended to $D$ dimensional Einstein gravity \cite{Adami:2021nnf}. In \cite{Adami:2021kvx}, we established that the boundary degrees of freedom, associated with a generic codimension one null surface in $D$ dimensional pure Einstein gravity, naturally lend themselves to a thermodynamic description. Furthermore, we considered the causal (timelike or null) boundaries for $2d$ and $3d$ gravity theories in \cite{Adami:2022ktn}, and presented its hydrodynamic description in \cite{Adami:2023fbm}.

\paragraph{Main results.} We extend the RT solution to accommodate the presence of a codimension 1 boundary which we typically choose to be a causal hypersurface in a background with generic cosmological constant $\Lambda$ and allow for the wavefront to be a smooth compact $2d$ surface of arbitrary topology. Our solution space beside the usual time dependent mass and RT scalar mode which describes the time dependence and shape of the RT wavefront and the genus of the wavefront is described by two scalar boundary (soft) modes and a vector field that indicates the local angular velocity of causal hypersurfaces. The two boundary scalar modes are described by two generic functions on the codimension 1 boundary and the vector field which is defined on a codimension 1 surface, satisfies the conformal Killing equations (CKE) over the wavefront. 

After constructing the solution space we analyze boundary symmetries that consist of a supertranslation in time (which only depends on time), codimension 1 radial supertranslation and radial Weyl scalings, and codimension 1 superrotations which satisfy the CKE over the wavefront. 

Next, we study the on-shell symplectic form over the solution phase space and then the surface charges associated with the boundary symmetries. The symplectic form consists of 2 parts: ``thermodynamic'', ``codimension 2'' (boundary symplectic form). The bulk RT mode does not appear in the symplectic form, as discussed in \cite{Adami:2023wbe}. We find that only particular combinations, contingent upon the chosen slicing \cite{Adami:2020ugu, Adami:2021nnf, Adami:2022ktn}, result in non-zero charges for the given boundary symmetries. We demonstrate the existence of a Heisenberg slicing, wherein the surface charge algebra adopts the structure of a Heisenberg algebra, with the time-independent (ADM) mass serving as a Casimir of the algebra. Furthermore, we study charges associated with superrotation symmetry generators and discuss that there is no independent charge (other than the charges appearing in the Heisenberg algebra) associated with them.

{Lastly, we study the generalized RT solution within the fluid/gravity correspondence framework \footnote{ For a review of prior efforts in this direction with various methodologies see \cite{Ciambelli:2018wre, Miskovic:2023zfz, Campoleoni:2023fug, Ciambelli:2018xat, Ciambelli:2020ftk, Campoleoni:2022wmf, Mittal:2022ywl}.}. Considering the on-shell RT symplectic form and surface charges, we construct the RT energy-momentum tensor for both asymptotically AdS and flat spacetimes. In the context of the AdS case, our findings extend the analysis presented in \cite{BernardideFreitas:2014eoi, Ciambelli:2017wou}. For the flat case, we construct a null energy-momentum tensor, utilizing the Carrollian structure of the null infinity. We then consider its conservation through a Carroll-Affine connection. The conservation of this energy-momentum tensor results in the RT equation. Interestingly, we demonstrate that the RT equation can be written in terms of a current conservation equation, in addition to its tensor representation. This is applicable for both AdS and flat cases.}

\paragraph{Outline of the paper.} In section \ref{sec:solution}, we systematically construct the most general solution space within Einstein-$\Lambda$ theory, excluding tensorial bulk modes, while rigorously examining the symmetries inherent to the solution space.
Section \ref{sec:symp-structure} is dedicated to a comprehensive analysis of the symplectic structure of the solution space. Proceeding to section \ref{sec:symp-form}, we undertake an in-depth discussion concerning the symplectic form and surface charges, with a specific focus on compact wavefront topologies. In sections \ref{sec:BEMT} and \ref{T-flat}, we introduce the boundary energy-momentum tensor corresponding to the thermodynamical/hydrodynamical symplectic structure and explore the fluid description of our extended RT solution space for asymptotically Anti-de Sitter (AdS) and flat (Carrollian) spacetimes. Finally, Section \ref{sec:discusion} contains a brief summary, discussion, and prospects for future research endeavors. In appendix \ref{sec:topologies}, we have gathered some details of the symplectic form for various topologies.

\section{Constructing solution space}\label{sec:solution}
Consider the line element in the Gaussian-like coordinate system in $4d$ \begin{equation}\label{metric-GTC}
    \d s^2= g_{\mu\nu}\d x^\mu \d x^\nu= -V \d v^2 + 2 \eta \d v \d r + g_{AB} \left( \d x^A + U^A \d v \right) \left( \d x^B + U^B \d v \right)\, ,
\end{equation}
where $v$ is advanced time coordinate, $r$ is the radial coordinate and $x^A\, , A=1,2,$ are coordinates on a compact two dimensional surface $\Sigma_g$. We fix the coordinate such that $\eta$ does not depend on radial coordinate while $V$, $U^A$ are generic functions on the spacetime. We study a class of geometries with the following properties:
\begin{enumerate}
    \item They admit a shear and twist free null geodesic congruence.
    In the adopted coordinate system, 
    the given null congruence is along vector field $\partial_r$. The radial coordinate $r$ is chosen such that it is the affine parameter along the null congruence. 
    \item The null congruence in general has a non-vanishing expansion.
    \item 
    We take $r=0$ surface as the presumed boundary in the spacetime where we define our boundary data and boundary dynamics.
    Note that constant $r$ surfaces can be timelike, null, or spacelike depending on the sign of $V$.
    \item The codimension two transverse surface $\Sigma_g$ is conformally related to a fixed metric $q_{AB}$ on the transverse surface $\Sigma$. We suppose that $q_{AB}$ does not depend on advanced time $v$ and radial coordinate $r$ and that
\begin{equation}\label{transverse-metric}
    g_{AB}=\Phi^2\,q_{AB} \, , 
\end{equation}
where $\Phi$ is a generic function on spacetime. As we will see below, the above is a consequence of requiring the null congruence to be shear-free. 
\item We assume $\Sigma$ to be a orientable, smooth and compact $2d$ surface with a generic genus $\mathrm{g}$.
\end{enumerate}
Throughout the paper, we will lower and raise Capital Latin letters by $q_{AB}$ and its inverse $q^{AB}$ respectively.

\paragraph{Null geometric quantities.} 

As pointed out the class of geometries of our interest admits a shear and twist free geodesic null congruence. To make this more explicit, consider the following two null vector fields
\begin{equation}\label{null-basis'}
\begin{split}
    l_{\mu} \d x^{\mu} &= -\frac{1}{2} V \d v  + \eta \d r \, , \qquad\qquad
    n_{\mu} \d x^{\mu} = -\d v \, , \\
    l^{\mu}\partial_{\mu} &=\partial_{v}+\frac{V}{2\eta}\partial_{r}-U^A\partial_{A}\, , \hspace{1 cm} n^{\mu}\partial_{\mu}=-\frac{1}{\eta}\partial_{r}\, ,
\end{split}
\end{equation}
with normalization condition
$ l\cdot n=-1$. Furthermore, we define the induced metric on the transverse surface $\Sigma_{g}$ as 
\begin{equation}
    \text{h}_{\mu\nu}=g_{\mu\nu}+l_{\mu}n_{\nu}+l_{\nu}n_{\mu}\, .
\end{equation}
Independent components of the covariant derivative of these two vector fields $l,n$, yield useful geometric quantities.  The expansions $\theta_l,\theta_n$, and the non-affinity parameter $\kappa$ are
\begin{subequations}\label{kappa-thetal-thetan}
\begin{align}
     \kappa &:=-l^{\alpha} n^{\beta}\nabla_{\alpha}l_{\beta} = \frac{\text{D}_v\eta}{\eta}+\frac{\partial_r V}{2\eta}, \label{kappa}\\ 
     \theta_l&:=\text{h}_{\alpha\beta}\nabla^{\alpha} l^\beta=2\frac{\text{D}_{v}\Phi}{\Phi}+
    \frac{V}{\eta }\frac{\partial_{r}\Phi}{\Phi}\, ,\label{theta-l}\\
  \theta_n &:=\text{h}_{\alpha\beta}\nabla^{\alpha} n^\beta=-\frac{2}{\eta}\frac{\partial_{r}\Phi}{\Phi}
\, .\label{theta-n}
    \end{align}
\end{subequations}
where we use $\text{D}_{v}:=\partial_{v}-\mathcal{L}_{U}$. We also note that the non-affinity of $n^\mu$ vanishes. The deviation tensor of the null vector field $l^\mu$ can be decomposed as
\begin{equation}\label{}
       B^{l} _{\mu \nu} :=  \text{h}^{}_{\mu \alpha} \text{h}^{}_{\nu \beta }\nabla^{\beta} l^{\alpha} = \frac{1}{2}   \theta_l \, \text{h}_{\mu \nu} + N^{l}_{\mu \nu}\, , 
\end{equation}
where $N^{l}_{\mu \nu}$
is the shear tensor. Transverse components of $ N^{l}_{\mu\nu}$ are given by
\begin{equation}\label{NAB-l}
    N^{l}_{AB}
    =\frac{1}{2}\, \Phi^2\, \text{D}_{v}q_{AB}=-\frac{1}{2}\, \Phi^2\left(D_{A}U_{B}+D_{B}U_{A}-q_{AB}D_{C}U^{C}\right)\, ,
\end{equation}
where \eqref{transverse-metric} is used.  As we see $N^{l}_{AB}=0$ if $\text{D}_{v}q_{AB}=0$. 
We can apply the same procedure to a null vector field $n^{\mu}$ and introduce a second-rank tensor $B^{n}_{\mu \nu} := \text{h}_{\mu \alpha} \text{h}_{\nu \beta }\nabla^{\beta} n^{\alpha}$. When we impose the condition that the null vector field $n$ is shear-free ($N^{n}_{\mu\nu}=0$), this requirement leads to \eqref{transverse-metric}. Additionally, it becomes evident that $n$ is twist-free as $\nabla_{[\mu} n_{\nu]}=0$.

\subsection{Various wavefront topologies} 

Codimension 2 surface $\Sigma$, which is the wavefront of the RT gravitational wave, can admit various (compact) topologies $S^2, T^2$ or a handle body object of genus $\mathrm{g}$ for different values of cosmological constant $\Lambda$. To see this, it is handy to choose $x^A$ to be the complex coordinates $z,\bar z$ such that the metric has a conformally flat form
\begin{equation}\label{metric-001}
    q_{AB} \d{}x^A\d{}x^B :=2  q_{z\bar{z}} \d{}z \d{}{\bar z}=\frac{2}{(1+\varepsilon\, z\bar{z})^{2}}\ \d{}z \d{}{\bar z}\, ,
\end{equation}
with $\varepsilon=0,\pm 1$ correspond to $2d$ curvature spaces, $R[q]=2\varepsilon$. All solutions that we study are parametrized by two parameters $\Lambda$ and $\varepsilon$.

The genus of $2d$ surface $\mathrm{g}$ which is also defined as 
\begin{equation}\label{genus}
    \begin{split}
        \mathrm{g}=\ 1 - \frac{1}{8\pi}\int \d{}^2x \, \sqrt{q}\,  \mathrm{R}[q]
        =\ 1+\frac{i}{4\pi}\int \d z \d{} \bar{z} \, \partial \bar{\partial} \ln{q_{z\bar{z}}}\, . 
    \end{split}
\end{equation}
We note that for metric \eqref{metric-001} $\sqrt{q}=i\,  q_{z\bar{z}} $.

\paragraph{$\bullet$ $\varepsilon=1$, spherical topology:} In this case, $q_{AB}$ describes a unit sphere and we take $z=\tan{(\theta/2)} \, e^{i\phi}$, where $0 \leq \theta \leq \pi$, $0 \leq \phi \leq 2\pi$. The area element is
\begin{equation}
    \frac{2  i}{(1+ z \bar{z})^2} \, \d z \wedge \d{} \bar{z}= \sin{\theta} \d \theta \wedge \d \phi\, .
\end{equation}
In this case,
\begin{equation}
       \oint \d z \d{}\bar{z} \,\sqrt{q}\,z^{m-1} \bar z^{n-1}  = 4\pi \, \delta_{m,n} \, I(m)
\end{equation}
where
\begin{equation}
         I(m):=\frac{1}{2} \int_{-1}^1  \d \mu \,  \left(\frac{1-\mu}{1+\mu}\right)^{m-1}\, , \qquad \mu= \cos{\theta}\, .
\end{equation}
It is easy to see that $I(m)$ diverges for all $m \neq 1$ while $I(1)=1$. Note also that $I(-m+2)=I(m)$.

\paragraph{$\bullet$ $\varepsilon=0$, toroidal topology:} This case corresponds to torus and we choose $z= (\vartheta + \tau \varphi)/2 $ where $\tau$ is a constant complex number (complex structure of the torus) and $0\leq \vartheta,\varphi < 2\pi$. The area element is 
\begin{equation}
    2 i \d z \wedge \d{} \bar{z}=  \mathrm{Im}(\tau)\,  \d \vartheta \wedge \d \varphi \, ,
\qquad \Rightarrow \qquad
   2 i \int  \d z \wedge \d{} \bar{z}= 4 \pi^2\, \mathrm{Im}(\tau) \, .
\end{equation}
On a torus, we shall expand a generic function of $\vartheta$ and $\varphi$ as $\chi(\vartheta,\varphi)= \sum_{n,m} \chi_{m,n} e^{i ( m \vartheta+ n \varphi)}$.
One can simply infer that 
\begin{equation}\label{Torus-identity}
       \oint \d z \d{}\bar{z} \,\sqrt{q}\,e^{i ( m \vartheta+ n \varphi)}  = 4\pi^2 \, \mathrm{Im}(\tau)\, \delta_{m,0} \delta_{n,0} 
\end{equation}

\paragraph{$\bullet$ $\varepsilon=-1$, $\mathrm{g}>1$ topology:} In this case we deal with a hyperboloid $\mathbb{H}^2$. $\mathbb{H}^2$ has uniform negative curvature, is non-compact and has isometry $so(2,1)$. One makes it a compact space through orbifolding it by freely-acting discrete subgroups of its isometry group. Depending on the choice of the subgroup we obtain a smooth compact manifold with genus $\mathrm{g}>1$. The action of this orbifolding can be reflected in the range of coordinates parameterizing the $H^2$, $z=\tanh{(\psi/2)}\, e^{i\phi}$, where $0 \leq \psi < \cosh^{-1}{(2\mathrm{g}-1)}$ and $\phi\in [0,2\pi]$. 
The area element is
\begin{equation}
    \frac{2  i}{(1- z \bar{z})^2} \, \d z \wedge \d{} \bar{z}= \sinh{\psi} \d \psi \wedge \d \phi\, .
\end{equation}
In this case
\begin{equation}
         I(m;-1):=\frac{1}{2} \int^{2\mathrm{g}-1}_{1}  \d \mu \,  \left(\frac{\mu-1}{\mu+1}\right)^{m-1}\, , \qquad \mu=\cosh\psi\, .
\end{equation}
A distinct way of realizing the periodic boundary conditions imposed by the given discrete group is upon cutting the surface of genus g which will result in a polygon of $4\mathrm{g}$ edges. Then we can triangulate this polygon to get the area $A$ of the surface. If its curvature $\varepsilon$ is constant we can use \eqref{genus} to obtain $A  = 4\pi (\mathrm{g} -1)$.

\subsection{Einstein field equations}\label{sec:EoM}

We now impose  Einstein equations $G_{\mu\nu} + \Lambda g_{\mu\nu}=0$, where $G_{\mu\nu}$ denotes the Einstein tensor and $\Lambda$ is the cosmological constant, on the metric ansatz \eqref{metric-GTC}. The $rr$--component of Einstein equations fixes the $r$ dependence of $\Phi$ as
\begin{equation}
    \Phi:=\sqrt{\Omega} +r\, \eta\, \lambda \, .
\end{equation}
where $\Omega, \eta, \lambda$ are arbitrary functions of $v,x^A$. With the above, $\theta_n$, the expansion of the null vector field $n$,  is \eqref{theta-n}
\begin{equation}
    \theta_n=-\frac2\eta \frac{\partial_r\Phi}{\Phi}=-\frac{2\lambda}{\Phi}\, .\label{theta-n-1}
\end{equation}
As we see $\theta_n$ is generically non-zero and can be zero only if $\lambda=0$. In our analysis here we always assume $\lambda\neq 0$ and exclude the family of $pp$-waves (Kundt class \cite{kundtpp-wave}). Moreover, as we see the expansion at large $r$ goes as $1/r$. At our presumed boundary at $r=0$, $\theta_n=\lambda/\Omega$. 

The $rA$- and $AB$- components of the equations of motion lead to
\begin{subequations}\label{RT-solution}
    \begin{align}
        & U^{A}=\mathcal{U}^A+\frac{D^{A}\eta}{\eta \, \lambda}\, \Phi^{-1}+\eta \, D^{A}e^{-\Pi/2}\, \Phi^{-2} \, , \label{U-expression}\\
        & V= -\frac{\Lambda}{3\, \lambda^2} \, \Phi^{2}-\frac{1}{\lambda} \left( \mathcal{D}_v \Pi + {\frac{\mathcal{D}_v (\Omega\sqrt{q})}{\Omega\sqrt{q}}}\right) \, \Phi+\mathcal{V} +\mathcal{X} \, \Phi^{-1}+\frac{\Omega}{4\, \lambda^2}\, D_{A}\Pi \, D^{A}\Pi \, \Phi^{-2}\, ,\label{V-expression}
    \end{align}
\end{subequations}
where
\begin{equation}\label{Pi-def}
    \Pi:=\ln{\left(\frac{\eta^2\lambda^2}{\Omega}\right)}\, ,
\end{equation}
and
\begin{equation}
 \mathcal{V}:=\frac{\sqrt{\Omega}}{\lambda}\, \mathcal{D}_{v}\Pi+\frac{D_{A}\eta \, D^A \eta}{\eta^2 \, \lambda^2}-\frac{D^2 \lambda}{\lambda^3}+\frac{D_{A} \lambda \, D^A  \lambda}{\lambda^4}+\frac{\mathrm{R}[q]}{2\, \lambda^2}\, .
\end{equation}
In the above equations $\mathrm{R}[q]$ and $D_{A}$ respectively denote Ricci scalar and covariant derivative with respect to $q_{AB}$ and $D^2:=D^AD_A$. We also define the operator $\mathcal{D}_{v}$ with $\mathcal{D}_{v}:= \partial_{v}-{\mathcal{L}}_{\mathcal{U}}$
where ${\mathcal{L}}_{\mathcal{U}}$ is the usual Lie derivative. The $vA$-components specify $\mathcal{X}$ as
\begin{equation}
    \mathcal{X}= -2 \, m \, G \, \lambda -\frac{\sqrt{\Omega}}{\eta\, \lambda^2}\, D_A \Pi\, D^A \eta\, ,
\end{equation}
where $m=m(v)$ does not depend on coordinates on the transverse surface.

The $AB$-component of equations of motion also yields the conformal Killing equation (CKE) for $\mathcal{U}^A=\mathcal{U}^A (v, x^A)$, 
\begin{equation}\label{EoM-calU}
D_{A}\mathcal{U}_{B}+D_{B}\mathcal{U}_{A} - D_{C}\mathcal{U}^Cq_{AB}=0\, .
\end{equation}
The above using \eqref{U-expression} implies $\lim_{r \to \infty}(\Phi^{-2}N_{AB}^{l})=-\frac{1}{2}(D_{A}\mathcal{U}_{B}+D_{B}\mathcal{U}_{A} - D_{C}\mathcal{U}^Cq_{AB})=0$. That is, the CKE \eqref{EoM-calU} is equivalent to the vanishing of the news tensor at infinity.

The $vv$-components of Einstein's equations implies 
\begin{equation}\label{RT-eq}
    \partial_{v}(m {\hat{\lambda}^{3}}) 
    - \frac32\,{\hat{\lambda}} \,  \hat{D}_{A}(m \hat{\lambda}^{2}\mathcal{U}^A)- \frac{q}{8G}\ {\hat{\lambda}}\, \hat{D}^{2}\left(\hat{\lambda}^{-2}\hat{D}^{2}\ln{\hat{\lambda}^2}
    \right)=0\, ,
\end{equation}
where $\hat{\lambda}=\lambda q^{1/4}$ and $\hat{D}_A$ is metric connection compatible with $\hat{q}_{AB}\d x^A \d x^B =2 \d z \d{\bar{z}}$. We also define $\hat{\Omega}= \sqrt{q}\Omega$ for later convenience. 
The above equation is called \textit{Robinson-Trautman equation} (RT equation) and is the only dynamical equation left on the given spacetime. It is worth stressing that the RT equation does not depend on the cosmological constant and $\Lambda$ appears only in the expression for $V$ \eqref{V-expression}.

\subsection{More on solutions to field equations}

\paragraph{Solutions to $2d$ conformal Killing equation \eqref{EoM-calU}.} In the $4d$ case we consider here, $A=1,2$ and solutions to \eqref{EoM-calU}  yield the $2d$ conformal algebra. It has an infinite tower of solutions that their Lie bracket form $Witt\oplus Witt$ algebra.  To make the above explicit recall that any $2d$ metric can be (locally) brought to a conformally flat form. In this coordinate system, the CKE \eqref{EoM-calU}, namely $2 q_{z\bar{z}}\partial \mathcal{U}^{\bar{z}}=0$ and $2 q_{z\bar{z}}\bar{\partial} \mathcal{U}^{{z}}=0$, can be solved
\begin{equation}\label{CKE-U}
    \mathcal{U}^{A}\partial_{A}=\mathcal{U}^{z}(v,z)\partial_{z}+ \mathcal{U}^{\bar{z}}(v,\bar{z})\partial_{\bar{z}}\, .
\end{equation}
Reality of ${\cal U}$ implies $\overline{\mathcal{U}^{z}(v,z)}=\mathcal{U}^{\bar{z}}(v,\bar{z})$. Assuming that $\mathcal{U}^{z}(v,z)$ is a meromorphic function on $z$-plane it admits a Laurent expansion,
\begin{equation}\label{U-zbarU}
    \mathcal{U}^{z}(v,z)=-\sum_{n\in\mathbb{Z}} {\cal U}_n(v)\, z^{-n+1}, \qquad {\cal U}_n(v)=-\frac{1}{2\pi i} \oint \d{}z\  \mathcal{U}^{z}(v,z) z^{n-2}\, .
\end{equation}
To exclude poles at $z=0$ and $z=\infty$, we shall consider only six modes, namely $\{\mathcal{U}_{-1}(v),\mathcal{U}_{0}(v),\mathcal{U}_{1}(v) \}$.

\paragraph{Solving RT equation \eqref{RT-eq}.} Integrating \eqref{RT-eq} over the complex plane (compact 2-surface), yields the equation for $m$,  $\partial_{v}(m\, \mathcal{R}^3)=0$ where 
\begin{equation}\label{R2-def}
    {\cal R}(v)^2:= 
    \frac{1}{4\pi} \oint \d{}z \d{} \bar{z} \, \sqrt{q}\, \lambda^2= \frac{1}{16\pi} \oint \sqrt{{}^{\text{\tiny(2)}}g} \, \theta_n^2
    \, ,
\end{equation}
where $\sqrt{{}^{\text{\tiny(2)}}g}=\sqrt{q}\ \Phi^2$ and
in the last equality, we used \eqref{theta-n-1}.
We discuss later the conditions for which the right-hand side of the above is finite. In other words,
\begin{equation}\label{mR3-EoM-2}
m= \frac{{\cal M}_\circ}{\mathcal{R}^3}\, ,
\end{equation}
with ${\cal M}_\circ$ being an integration constant. The above suggests (see discussions in section \ref{sec:boundary-symmetry}) defining a new time coordinate $\beta$,
\begin{equation}\label{beta-v}
    \beta (v) := \int^v \mathcal{R}^{-1} \d v\,.
\end{equation}
Adopting $\beta$ as time coordinate, one should redefine  $\tilde{\mathcal{U}}^A =  \mathcal{R} \, \mathcal{U}^A$.

As the next step, we plug back the above $m(v)$ into \eqref{RT-eq} to obtain an equation for  $\hat{\lambda}^2$. This is a nonlinear but first order equation for $\tilde{\lambda}=\hat{\lambda} /\mathcal{R}$ which takes a simple form
\begin{equation}\label{RT-eq-2}
  3 {\cal M}_{\circ} G\,  {\cal D}_{\beta} \tilde{\lambda}^2 
   =\partial\bar\partial \left(\tilde\lambda^{-2} \partial\bar\partial \ln \tilde\lambda^2\right) \qquad \Longrightarrow \qquad \tilde{\lambda}=\tilde{\lambda}(\beta; {\cal U}^A; \tilde{\lambda}_b)\, ,
\end{equation}
where ${\cal D}_{\beta} :=\partial_\beta - \mathcal{L}_{\tilde{\mathcal{U}}}$,
and $\tilde{\lambda}_b:=\tilde{\lambda}_b(z,\bar z)$ are integration constants which may be viewed as $\tilde{\lambda}_b(z,\bar z)=\tilde{\lambda}(v=v_b; z,\bar z)$. That is, \eqref{RT-eq} specifies $\tilde{\lambda}(v; z,\bar z)$ up to a codimension 2 function. RT equation \eqref{RT-eq-2} is compatible with various wavefront topologies. So, along with the shape function $\tilde{\lambda}_b(z,\bar z)$ one should also specify topology of the wavefront and in particular its genus $\mathrm{g}$. 

To summarize, the solution space is specified through codimension 1 functions $\Omega(v; z,\bar z), \eta(v; z,\bar z)$; codimension 2 functions ${\cal U}(v;z), {\bar{\cal U}}(v;{\bar z})$, $\tilde\lambda_b(z,\bar z)$;  ${\cal R}$ as function of only $v$ and finally the constant ${\cal M}_{\circ}$ and the genus of wavefront $\mathrm{g}$. We will address the question that which of these functions are physical and which are pure gauge and can be removed upon proper diffeomorphisms in section \ref{sec:symp-form} where we analyze the surface/boundary charges.

\subsection{Singularity, past and future  horizons}

Singularity and causal structure, past and future horizon, and global structure of RT solutions have been discussed in the literature, including \cite{chrusciel1992global, Griffiths:2002gm, Bicak:1997ne, Bicak:1999ha, Bicak:1999hb, Chrusciel:1992tj, 1992RSPSA.436..299C, Svitek:2011zz}. In what follows we review and extend these results to our extended RT solution space. 

\paragraph{Global structure of the RT solution.}
The RT solution in equation \eqref{RT-solution} contains the following free data \footnote{We will show that only the zero mode of $\beta(v)$, namely $\beta(v_b)$, appears in the symplectic form (see equation \eqref{symp-form-reg}).}
\begin{equation}
   \text{Free data:} \qquad \{\mathcal{M}_\circ, \mathrm{g} , \beta(v_{b}),\lambda(v_{b},x^A),\Omega(v,x^A),\eta(v,x^A), \mathcal{U}^{A}(v,x^B),q_{AB}(x^A)\}\, .
\end{equation}
We note that the topology of the wavefront, i.e. its genus $\mathrm{g}$, and the global and causal structure of the RT geometry do not change under regular diffeomorphisms. In this regard, to discuss the global structure we use local diffeomorphisms to fix
\begin{equation}\label{gauge-fixed-soln}
    \begin{split}
        &  \Omega(v,x^A)=0\, , \qquad \eta(v,x^A)= 1\, , \qquad q_{AB}=2d\ \text{constant curvature metric}\, .
    \end{split}
\end{equation}
With the above, $\Phi=r\lambda$.  One can use the regular diffeomorphisms to turn off the global part of \eqref{CKE-U} namely $\{\mathcal{U}_{-1}(v),\mathcal{U}_{0}(v),\mathcal{U}_{1}(v) \}$ and their complex conjugates. The non-global part of $\mathcal{U}^{A}$ cannot be eliminated using regular diffeomorphisms without introducing poles on the wavefront $\Sigma$. However, $\mathcal{U}^{A}$ do not enter into the causal and singularity structure of our solutions and we ignore it in the discussions below. 

The global structure of the RT solution hinges on three parameters: $\{\Lambda, \mathrm{g}, \mathcal{M}_\circ\}$. The cosmological constant $\Lambda$ can assume three distinct cases: the asymptotically flat case where $\Lambda=0$, the asymptotically AdS case with $\Lambda<0$, and the asymptotically dS case where $\Lambda>0$. The second parameter, $\mathrm{g}$, denotes the genus of the wavefront, $\mathrm{g}=0,1$ respectively correspond to topologically spherical and toroidal wavefronts and $\mathrm{g}>1$ cases correspond to negative curvature wavefronts. The third parameter, $\mathcal{M}_\circ$, corresponds to the ADM mass and can be positive ($\mathcal{M}_\circ>0$), zero ($\mathcal{M}_\circ=0$), or negative ($\mathcal{M}_\circ<0$). Given these possibilities, we can conclude that there are 3×3×3=27 distinct global structures for RT spacetimes.

Here we only discuss the global structures of positive mass cases ($\mathcal{M}_\circ>0$):\footnote{For negative mass cases, see \cite{1992RSPSA.436..299C}.}
\begin{itemize}
\item Singularity: The Kretschmann scalar ($R_{\mu\nu\alpha\beta}R^{\mu\nu\alpha\beta}$) diverges at $r=0$ as $r^{-6}$. Hence $r=0$ is a curvature singularity.\footnote{Note in general the singularity is at $\Phi=0$, which with the choice in \eqref{gauge-fixed-metric} it corresponds to $r=0$.} In all cases, it is a naked singularity.  Intriguingly, even in the presence of such a naked singularity, the evolution of the metric remains unique \cite{chrusciel1992global}.
\item Incompleteness of $\mathcal{J}^{+}$: The common feature of RT solutions (even in the $\mathcal{M}_\circ<0$ case) is that the future infinity $\mathcal{J}^{+}$ is not complete \cite{1992CMaPh.147..137C, chrusciel1992non}. This fact in the asymptotically flat case leads to the absence of spatial infinity $i^{0}$. This result also extends to $\Lambda\neq 0$ cases \cite{Griffiths:2002gm, Bicak:1995vc, Bicak:1999ha, Bicak:1997ne}. 
\item Horizon: To solve the RT equation we give the initial data on a null hypersurface (i.e. $v=v_b$) then the RT equation evolves this data. Finally, at large times we end up with a Schwarzschild-like horizon \cite{chrusciel1992global}. 
\item The non-smoothness of event horizons: The RT horizon exhibits an intriguing characteristic; its smoothness is finite and varies in degree. This variation is directly influenced by the cosmological constant. As it increases, so does the smoothness of the horizon. For a positive cosmological constant, there exists a critical threshold, denoted as $\Lambda_{c}=\frac{1}{9\mathcal{M}_{\circ}^2}$. Beyond this value, the horizon achieves infinite smoothness. However, it’s important to note that despite this infinite smoothness, the horizon is not analytical \cite{Bicak:1997ne}.
\end{itemize}
The algebraically special class of Robinson–Trautman spacetimes can be classified in Petrov types II, D, III, and N, while cosmological constant $\Lambda$ can be included easily in the discussion. Based on the components of the Weyl tensor, it can be demonstrated that RT spacetimes for both cases $\mathcal{M}_{\circ} = 0$ and $ \mathcal{M}_{\circ} \neq 0$ have curvature singularity at $r = 0$ \cite{Stephani:2003tm, Griffiths:2009dfa} and become conformally flat, dS or AdS at the limit $r \to \infty$ according to the value of $\Lambda$. RT spacetimes for $\mathcal{M}_{\circ} \neq 0$ have algebraic types II or D, which contain the Schwarzschild and the C-metric solutions or expanding gravitational waves which decay to Schwarzschild at the asymptotic $v \to \infty$ \cite{1992CMaPh.147..137C, chrusciel1992non}. For case $\mathcal{M}_{\circ} = 0$, authors of \cite{Griffiths:2002gm} give an interpretation for the type N of RT spacetimes which describes pure expanding gravitational waves as sandwich waves for all values of the $\Lambda$ and $\mathrm{g}$.

As a final point in this subsection, we note that in this paper we restrict ourselves to $\lambda \neq 0$. From equation \eqref{theta-n-1}, this corresponds to expanding horizons. In other words, we exclude the family of $pp$-wave solutions.

\subsection{Large $r$ asymptotic behavior} 
The class of solutions we are considering  $r$-dependence  appears only through the function $\Phi$. Therefore, in the large $r$, $\Phi\simeq \eta\lambda r$. Recalling the form of $V$ \eqref{V-expression}, the leading behavior of $V$ depends on $\Lambda$. Nonetheless, RT equation \eqref{RT-eq} is $\Lambda$ independent and the RT equation has solutions for different choices of the topology of the wavefront $\Sigma$. As mentioned below \eqref{CKE-U}, the $r^2$ order in $N_{AB}^l$ \eqref{NAB-l} vanishes and the next to leading term in the large $r$ is
\begin{equation}\label{NAB-l-leading-r}
    N_{AB}^l= r \left[D_{\langle A}\ \eta D_{B \rangle}\ln(\eta\lambda)^2-D_{\langle A}D_{B \rangle}\eta\right] + {\cal O}(1),
\end{equation}
for both asymptotic flat or AdS cases. In the above $X_{\langle A}Y_{B \rangle}=\frac12(X_AY_B+ X_B Y_A-q_{AB} X^C Y_C)$ is the traceless symmetric product. As we see the leading term in \eqref{NAB-l-leading-r} is completely specified through derivatives of $\eta$ and $\lambda$, which as we will see below, are the two functions appearing in the asymptotic boundary metric. 

\paragraph{Asymptotic AdS$_4$ case.} For  $\Lambda =-3/\ell^2 <0$, we have 
\begin{equation}\label{metric-asymptotic-Lambda}
    \d s^2=  -\frac{\eta^2}{\ell^2} r^2  \d v^2 + 2 \eta \d v \d r + r^2\ (\eta\lambda)^2 q_{AB}\left( \d x^A + {\cal U}^A \d v \right) \left( \d x^B + {\cal U}^B \d v \right)+ \cdots\, ,
\end{equation}
where $\cdots$ denote subleading terms. The first subleading term in $g_{vv}, g_{vA}, g_{AB}$ components of the metric is proportional to $r$ and involves the function $\Omega$. So, up to the first subleading term the above is an asymptotically AdS$_4$ space of radius $\ell$ and has 3 scalar functions of $v,x^A$, $\eta,\Omega, \lambda$ and a vector ${\cal U}^A(v,x^A)$.  $\eta$ may be absorbed into rescaling and a shift of $r$, i.e. by introducing $\rho=\eta r-\ell^2\partial_v\eta/\eta$ as the new radial coordinate, and in the leading part we remain with $\lambda, {\cal U}^A$ which are subject to 
\begin{equation}\label{lambda-calU-asymp}
D_{A}\mathcal{U}_{B}+D_{B}\mathcal{U}_{A} - D_{C}\mathcal{U}^Cq_{AB}=0\, , \qquad D^{2}\left[\frac{1}{{\lambda}^{2}}(D^{2}\ln{{\lambda}}- \frac12\mathrm{R}[{q}])\right]=0\, .
\end{equation}
The causal (conformal) boundary of the space is
\begin{equation}
    \d s^2_{b'dry}=  -{\ell^{-2}} \d v^2 + \lambda^2 q_{AB}\left( \d x^A + {\cal U}^A \d v \right) \left( \d x^B + {\cal U}^B \d v \right)\, .
\end{equation}
We note that the equation for $\lambda$ \eqref{lambda-calU-asymp} may be written as $\tilde{D}^2 (\tilde{D}^2 R[\tilde{q}])=0$ where $\tilde{q}_{AB}=\lambda^2 q_{AB}$ and $\tilde{D}_A$ is covariant derivative w.r.t the same metric. As we will argue below, the $x^A$ part of metric $\Sigma$, can be topologically an $S^2, T^2$ or a handle body object of genus $\mathrm{g}$.

\paragraph{Asymptotic flat case.} For $\Lambda=0$ case the asymptotic metric is of the form 
\begin{equation}\label{metric-asymptotic-flat}
    \d s^2=  {+}r\eta \mathcal{D}_v \hat\Pi \d v^2 + 2 \eta \d v \d r + r^2\ (\eta\lambda)^2 q_{AB}\left( \d x^A + {\cal U}^A \d v \right) \left( \d x^B + {\cal U}^B \d v \right)+ \cdots\, ,
\end{equation}
where $\hat\Pi=\ln(\eta^2\lambda^2\sqrt{q})$ and $\cdots$ denote subleading terms and ${\cal U}^A, \lambda$ are subject to \eqref{lambda-calU-asymp}. The constant large $\rho=\eta r \gg 1$ metric is of the form 
\begin{equation}\label{metric-asymptotic-flat-bdry}
    \d s^2_{b'dry}=  \rho^2\left[\frac{2}{\rho} \mathcal{D}_v \tilde\Pi \d v^2 + \lambda^2 q_{AB}\left( \d x^A + {\cal U}^A \d v \right) \left( \d x^B + {\cal U}^B \d v \right)\right], \qquad \rho=r\eta\, ,
\end{equation}
where $\tilde\Pi=\ln(\eta^2\hat{\lambda})$ and $\lambda$ is subject to \eqref{RT-eq}. We will argue below that the $x^A$ part of metric $\Sigma$ can be topologically an $S^2, T^2$ or a handle body object of genus $\mathrm{g}$. The above metric denotes a Carrollian geometry, see section \ref{sym-pot-LW} for a brief review, in which $1/\rho$ plays the role of speed of light $c$. We note that, unlike the asymptotic AdS case, $\eta$ still persists in the asymptotic metric through $\tilde\Pi$. 

For the $\Lambda>0$ case, we deal with an asymptotically de Sitter space and the large $r$ regime is behind the cosmological horizon whose radius is $\sqrt{\Lambda/3}$. For this case too, the $x^A$ part of metric $\Sigma$ can be topologically an $S^2, T^2$ or a handle body object of genus $\mathrm{g}$.

\paragraph{Comparison with BMS gauge and beyond.} The RT solution does not fit in the standard BMS falloff conditions \cite{Bondi:1962, Sachs:1962, Sachs:1962wk}. Various extensions of the BMS group by assuming different falloff conditions have been considered, e.g: Extended BMS \cite{Barnich:2010eb, Barnich:2009se, Barnich:2011ct}, Generalized BMS \cite{Campiglia:2015yka, Campiglia:2014yka}, Weyl BMS \cite{Freidel:2021fxf, Freidel:2021cjp, Freidel:2021qpz}, BMS in partial Bondi gauge \cite{Geiller:2022vto}, and  Logarithmic BMS \cite{Fuentealba:2022xsz, Fuentealba:2023syb} (for a review see \cite{Donnay:2023mrd, Ciambelli:2023bmn}). The reason is that in the usual BMS falloff conditions, the codimension 2 transverse surface has the following falloff $g_{AB}=r^2 \bar{q}_{AB}+r C_{AB}+\cdots$, where $\bar{q}_{AB}$ is the metric on unit sphere,  the shear tensor $C_{AB}$ is a symmetric traceless tensor, that is $C:=\bar{q}^{AB}C_{AB}=0$. The traceless condition arises from the determinant condition on transverse metric $\partial_{r}$det$(r^{-2}g_{AB})=0$.  By comparing the transverse metric \eqref{transverse-metric} with the mentioned falloff we recognize $\bar{q}_{AB}=\eta^2 \lambda^2 q_{AB}$ and $C_{AB}=2\sqrt{\Omega}\eta\lambda\, q_{AB}$. The point is that tensor $C_{AB}$ in our analysis is not taken to be traceless and its trace is given by $C=4\sqrt{\Omega}(\eta\lambda)^{-1}$. The reason is that we exploit different coordinate systems. See \cite{Adami:2023wbe} for more detailed analysis. 
In fact, the RT solution can be re-expressed in the Bondi coordinate system by performing a desired coordinate transformation, \emph{c.f.} section 5.2 in \cite{Adamo:2009vu}.

To cover the RT solution in the form \eqref{metric-GTC}, we need a weaker falloff such as the one discussed in \cite{Adami:2021nnf, Geiller:2022vto}. By relaxing the determinant condition, the first subleading term in the transverse metric is allowed to have a non-vanishing trace for $C_{AB}$. Strictly speaking, $\lambda$ in the solution space carries the information of the trace part of $C_{AB}$.

In summary, the solution phase space is parameterized by a codimension 3 function $m(v)$, and three codimension 1 scalars $\Omega\, , \eta \, , \lambda $, where $\lambda$ is subject to \eqref{RT-eq} and a codimension 1  vector $\mathcal{U}^A$ subject to \eqref{EoM-calU}. The constant $v, r$ surfaces $\Sigma$ can be a compact orientable surface with arbitrary genus $\mathrm{g}$. 
The Schwarzschild metric in Eddington–Finkelstein coordinates can be achieved by setting $\Omega=0$, $\lambda=\eta=1$, $\mathcal{U}^A=0$ and taking $q_{AB}$ to be the round metric on the unit sphere.

\section{Symplectic structure, general discussion}\label{sec:symp-structure}
The first-order variation of the Einstein-Hilbert action,
\begin{equation}\label{action}
    S=\int \d{}^4 x L[g]\, \, ,\, \qquad \text{with} \qquad L[g]=\frac{1}{16\pi G} \left[\sqrt{-g}  \left( R -2\Lambda \right)+\partial_{\mu} L_{\text{\tiny b}}^{\mu}[g]\right]\, ,
\end{equation}
give rise to two terms, the bulk and the boundary terms. The bulk term gives us equations of motion $G_{\mu\nu}+\Lambda g_{\mu\nu}=0$ and we refer to the boundary term as symplectic potential \cite{Lee:1990nz,Wald:1999wa} 
\begin{equation}\label{sym-pot}
    \bTh[g; \delta g]:=\int_{\mathcal{C}} \d{}^3 x_\mu \left( \Theta_{\text{\tiny LW }}^\mu [ g ; \delta g]+\delta  L_{\text{\tiny b}}^{\mu}[ g ] + \partial_{\nu} Y^{\mu \nu}[ g ; \delta g] \right)  \,.
\end{equation}
In the above, $L_{\text{\tiny b}}^{\mu}[ g ]$ is the boundary Lagrangian and it affects neither equations of motion nor symplectic form (and hence surface charge), $Y^{\mu \nu}[ g; \delta g]$ is a covariant skew-symmetric tensor constructed out of metric and its variation, and $\Theta^{\mu}_{_{\text{\tiny{LW}}}}[g; \delta g]$ is Lee-Wald's pre-symplectic potential,
\begin{equation}\label{}
    \Theta^{\mu}_{_{\text{\tiny{LW}}}} [g; \delta g]:=\frac{\sqrt{-g}}{8 \pi G} \nabla^{[\alpha} \left( g^{\mu ] \beta} \delta g_{\alpha \beta} \right)\,,
\end{equation}
where {$X^{[\alpha}Y^{\beta]} := \frac12 \left(X^{\alpha}Y^{\beta} -X^{\beta}Y^{\alpha}\right) $}. 
We refer to $Y^{\mu \nu}[ g; \delta g]$ as $Y$-freedom. 

\paragraph{Regularization of the symplectic potential.} The Lee-Wald pre-symplectic potential may lead to divergent terms at large $r$-limit. To make it finite, we need to remove the divergent terms exploiting two freedoms, namely $L_{\text{\tiny b}}^{\mu}$ and $Y^{\mu\nu}$, in the definition of the pre-symplectic potential \eqref{sym-pot}. To this end, let us consider Lee-Wald's pre-symplectic potential and assume that the boundary is located on a constant $r$ surface, then
\begin{equation}\label{sym-pot-LW}
    \begin{split}
        \bTh_{\text{\tiny LW }}[g; \delta g] &:= \ \int_{\mathcal{C}} \d v\d{}^2 x \, \Theta_{\text{\tiny LW }}^r  \\
       & =\ \frac{1}{16\pi G}\int_{\mathcal{C}_r} \d v\d{}^2 x \, \left\{-2\, G\, \sqrt{q}\, \lambda^2\, \delta m 
        +\mathcal{D}_{v}\left[\sqrt{q}\, \Phi^2\left(\delta {P}+\frac{\delta\Omega}{\Omega}\right)\right]\right\}\\
        &+  \frac{1}{16\pi G}\int_{\mathcal{C}_r} \d v\d{}^2 x \, \sqrt{q}\, D_A \left[ -2 \, \Phi^2\, \delta \mathcal{U}^A+\frac{\Phi}{\lambda}\, D^A\left(\delta {\mathrm{P}}+\frac{\delta\Omega}{\Omega}\right) -D^A \left( \frac{ \sqrt{\Omega}\, \delta \Pi}{2\, \lambda} \right)  \right]\\
        &+\frac{1}{16\pi G}\int_{\mathcal{C}_r} \d v\d{}^2 x \,\, \delta\left[\sqrt{q}\left(\frac{2\, \Lambda}{3\, \lambda}\, \Phi^3+4\, m \,G \, \lambda^2 \right)\right]\, ,
    \end{split}
\end{equation}
where
\begin{equation}\label{P-def}
    {\mathrm{P}}:= \ln{\left|\frac{\eta \, \lambda^2}{\Omega}\right|}\, .
\end{equation}

Consider the on-shell variation of the Einstein-Hilbert Lagrangian and write it suggestively \cite{McNees:2023tus}
\begin{equation}
    -\partial_{r}\Theta_{\text{\tiny LW }}^{r}\approx -\delta L_{\text{\tiny EH }}+\partial_{a}\Theta_{\text{\tiny LW }}^{a}\, .
\end{equation}
Integrating the above with respect to $r$, from $r=0$ to an arbitrary $r$, yields
\begin{equation}
    \Theta_{\text{\tiny reg. }}^r \approx \Theta_{\text{\tiny LW }}^r  +\delta  L_{\text{\tiny b}}^{r} + \partial_{a} Y^{r a}\, ,
\end{equation}
where $\Theta_{\text{\tiny reg. }}^r$ is the $r$-independent regularized symplectic potential,
\begin{subequations}
    \begin{align}
        &   L_{\text{\tiny b}}^{r}=-\int^{r}_0 \d r' \, L_{\text{\tiny EH }}(r')\Big|_{\text{\tiny{on-shell}}}=- \frac{\sqrt{q}}{16\pi G}\, \frac{2\Lambda}{3\lambda}\, \left(\Phi^3 -\Omega^{3/2}\right)\, , \label{L-r}\\
        & Y^{r v}=\int^{r}_0\d r'\, \Theta_{\text{\tiny LW }}^{v}(r')\Big|_{\text{\tiny{on-shell}}}= -\frac{\sqrt{q}}{16\pi G}\, \left(\Phi^2 -\Omega\right)\left(\delta {\mathrm{P}}+\frac{\delta\Omega}{\Omega}\right)\, , \label{Y-rv} \\
        & Y^{r A}=\int^{r}_0\d r'\, \Theta_{\text{\tiny LW }}^{A}(r')\Big|_{\text{\tiny{on-shell}}}= -\frac{\sqrt{q}}{16\pi G}\, \Bigl\{ \lambda^{-1}\left(\Phi-\sqrt{\Omega}\right)D^{A}\left(\delta {\mathrm{P}}+\frac{\delta\Omega}{\Omega}\right) \nonumber \\
        & \hspace{7 cm} - \bigg[2 \delta \mathcal{U}^A +\mathcal{U}^A \left(\delta {\mathrm{P}}+\frac{\delta\Omega}{\Omega}\right) \bigg] (\Phi^2-\Omega)\Bigr\}\, .\label{Y-rA}
    \end{align}
\end{subequations}
By setting $\Theta^\mu=\Theta_{\text{\tiny reg. }}^\mu$ and plugging the above in \eqref{sym-pot}, one finds
\begin{equation}\label{symplectic-pot'}
    \begin{split}
        \bTh_{\text{\tiny reg. }}&[g; \delta g]:= \ \int_{\mathcal{C}} \d v\d{}^2 x \, \Theta_{\text{\tiny reg. }}^r  \\
        =&\ -\frac{1}{8 \pi}\int \d v \d z \d{\bar{z}} \, \sqrt{q}\,\lambda^2 \delta m +\frac{1}{16\pi G}\int \d{}z\d{}\bar z \,  \hat{\Omega}\,  \delta {\mathrm{P}}  \\
       -& \frac{1}{16\pi G}\int \d v  \d{}z\d{}\bar z \,D_A  \left[ \hat{\Omega}\, (  \delta \mathcal{U}^A+  \mathcal{U}^A\, \delta  {\mathrm{P}})+\sqrt{q}\, \frac{\sqrt{\hat{\Omega}}}{\hat{\lambda}}\, D^A\left(\delta {\mathrm{P}}+\frac{\delta\hat{\Omega}}{\hat{\Omega}}\right) -\sqrt{q}\, D^A \left( \frac{ \sqrt{\hat{\Omega}}\, \delta \Pi}{2\, \hat{\lambda}} \right)\right]   \\
        +&\frac{1}{16\pi G}\int \d v \d{}z\d{}\bar z \, \delta\left[\frac{2\, \Lambda}{3\, \hat{\lambda}}\, \hat{\Omega}^\frac{3}{2}+4\, m \,G \, \hat{\lambda}^2 +\mathcal{D}_{v}\hat{\Omega}\right]\, .
    \end{split}
\end{equation}
The symplectic form can be obtained as the exterior derivative of symplectic potential on the phase space. In this regard, the symplectic form inferred from the symplectic potential \eqref{symplectic-pot'} can be split into the thermodynamical, boundary, and total derivative parts, i.e. 
\begin{equation}\label{symp-form-reg}
\begin{split}
    \bO_{\text{\tiny reg.}}[g; \delta g,\delta g] &= \bO^{\mathrm{thermo}}_{\text{\tiny reg.}}+ \bO^{\mathrm{b}}_{\text{\tiny reg.}}+\bO^{\mathrm{s}}_{\text{\tiny reg.}}\\ &= \delta \beta_\circ \wedge \delta \mathcal{M}_\circ+\frac{1}{16\pi G}\int \d{}z\d{}\bar z \,  \delta\hat{\Omega}\wedge  \delta {\mathrm{P}}+ \bO^{\mathrm{s}}_{\text{\tiny reg.}}
    \end{split}
\end{equation} 
where  $\beta_\circ= \beta(v_b)$ with $\beta(v)$ is defined in \eqref{beta-v} and 
\begin{subequations}
\begin{align}
&\bO^{\mathrm{thermo}}_{\text{\tiny reg.}}=\delta \beta_\circ \wedge \delta \mathcal{M}_\circ =-\frac{1}{8 \pi}\int \d v \d z \d{\bar{z}} \, \delta\hat{\lambda}^2 \wedge \delta m  \label{thermo-Omega}\\
   &\bO^{\mathrm{s}}_{\text{\tiny reg.}}= - \frac{1}{16\pi G}\int \d v \d{}z \d{\bar{z}}  \partial\left\{ \delta\hat{\Omega}\wedge (  \delta \mathcal{U}^z+  \mathcal{U}^z\, \delta  {\mathrm{P}})+ \delta \left(\frac{\sqrt{\hat{\Omega}}}{\hat{\lambda}}\right)\wedge \bar{\partial}\left(\delta {\mathrm{P}}+\frac{\delta\hat{\Omega}}{\hat{\Omega}}\right) -\bar{\partial}\left[ \delta\left( \frac{ \sqrt{\hat{\Omega}}}{2\, \hat{\lambda}} \right)\wedge  \delta \Pi\right]\right\}\nonumber\\
    \hspace{-0.45 cm} &\qquad +c.c.
\end{align}   
\end{subequations}
and we used \eqref{R2-def} and \eqref{mR3-EoM-2}. $\bO^{\mathrm{s}}_{\text{\tiny reg.}}$ is a codimension 1 integral over a total derivative. We emphasize that the codimension 1 integral of the right-hand-side of \eqref{thermo-Omega} is performed using the on-shell equations \eqref{R2-def} and \eqref{mR3-EoM-2}, and hence \eqref{symp-form-reg} is the on-shell symplectic.

The thermodynamical part is the one without time and space integrals. The boundary term $\bO^{\mathrm{b}}_{\text{\tiny reg.}}$, as in \cite{Adami:2023wbe}, involves a codimension 2 integral and $\bO^{\mathrm{s}}_{\text{\tiny reg.}}$ vanishes unless we allow for ``string-like'' sources. In this work, we focus on cases with no such sources and leave the discussion on string-like sources to future work. Compared to \cite{Adami:2023wbe}, our symplectic form does not involve a ``bulk'' term involving a codimension 1 spatial integral. The reason is in our case we have turned off the ``tensor modes'' (gravitational waves with non-compact wavefronts), $N_{AB}$ modes in the notation of \cite{Adami:2023wbe}, in our solution space. As discussed there, the RT modes, correspond to the kernel of the Carrollian solution space and hence drop out of the symplectic form once we turn off $N_{AB}$.   

We close this section with a comment on the RT mode $\lambda$. Despite the appearance in the solution space, $\lambda$ is absent in the on-shell symplectic form when the wavefront is smooth (and has no punctures). The symplectic potential \eqref{sym-pot-LW} involves $\lambda$ through its $\Phi$ dependence, recall that $\Phi=\sqrt{\Omega}+\eta \lambda r$. The $\lambda$ dependence, however, is removed when we remove $r$ dependence by our choice of ``regularizing'' $Y$-term, in particular \eqref{Y-rv} component. Therefore, $\lambda$ dependence could have been kept by other choices of $Y$-freedom. For all these cases, however, the symplectic form will have $r$-dependence and need not remain finite in the asymptotic large $r$ region.

\section{Symplectic form, boundary symmetries and surface charges for compact wavefront topologies}\label{sec:symp-form}

As mentioned, in the absence of punctures (sources) on the RT wavefront $\Sigma$, i.e. assuming $\Sigma$ to be a smooth compact handle body object with genus $\mathrm{g}$, $\bO^{\mathrm{s}}_{\text{\tiny reg.}}$ in \eqref{symp-form-reg} {and hence ${\cal U}^A$} drop out and we remain with thermodynamic and boundary terms in the on-shell symplectic form. In these cases, the information in the RT mode and its wavefront which is encoded in function $\lambda$,
except for the overall area of the wavefront encoded in $\beta$, drops out of the symplectic form. In other words, the boundary observer who is residing on the RT wavefront does not have access to or can't probe the details of the RT wavefront geometry. For this case, therefore, 
\begin{equation}\label{symplectic-form-com}
        \bO^{\text{\tiny on-shell}}_{\text{\tiny reg. }}[g; \delta g] = \delta \beta_\circ \wedge \delta \mathcal{M}_\circ  +\frac{1}{16\pi G}\oint \d{}z\d{}\bar z \,  \delta \hat{\Omega}\wedge  \delta {\mathrm{P}}\, .
\end{equation}
In what follows we first compute surface charges and their algebras and then to ensure the finiteness of symplectic form we shall discuss topologically different surfaces separately. 

\subsection{Boundary symmetry}\label{sec:boundary-symmetry}
The symmetry-generating vector field which preserves the form of the line element above is given by
\begin{equation}\label{BSG-1}
    \xi^{\mu}\partial_{\mu}=T\partial_{v}+\left(Z- r \, W+r \partial_vT\right)\partial_{r}+Y^{A}\partial_{A}\, ,
\end{equation}
where $T(v)$ is a codimension-three function parameterizing the $v$-dependence supertranslations. Also, $Z=Z(v,x^A)$ and $W=W(v,x^A)$ are arbitrary codimension-two functions. We refer to $Z$ and $W$ as supertranslation and superscaling in the radial direction. Finally, $Y^{A}=Y^A(v,x^B)$ represents the superrotations, and similarly to ${\cal U}^A(v,x^B)$, should  satisfy the CKE
\begin{equation}\label{CKT-Y}
    D_{A}Y_{B}+D_{B}Y_{A}-q_{AB}D_{C}Y^{C}=0\, .
\end{equation}
This condition is a direct consequence of $\delta q_{AB} = 0$. 
One can simply solve this equation for $Y^A$,
\begin{equation}\label{CKE-Y}
    Y^A \partial_{A}=Y^{z}(v,z)\partial_{z}+Y^{\bar{z}}(v,\bar{z})\partial_{\bar{z}}\, .
\end{equation}
With the Laurent expansion basis,
\begin{equation}\label{Y-v-z}
    Y^{A}\partial_A=\sum_{n\in\mathbb{Z}}  Y^z_n(v)\ L_n +\sum_{n\in\mathbb{Z}}  Y^{\bar{z}}_n(v)\ \bar L_n ,\qquad  L_n:= -z^{n+1}\partial_{z}, \quad \bar L_n:=-{\bar{z}}^{n+1} \,\partial_{{\bar{z}}}\, ,
\end{equation}
the Lie algebra of generators forms a $Witt\oplus Witt$ algebra, 
\begin{equation}
    [L_n, L_{m}]=(n-m)L_{n+m}\, , \quad [\bar L_{n}, \bar L_{m}]=(n-m)\bar L_{n+m}\, , \quad  [L_{n}, \bar L_{m}]=0\, .
\end{equation}
The global part of this algebra (the part which is invertible over  $\Sigma$) is an $sl(2,\mathbb{C})= sl(2,\mathbb{R})\oplus sl(2,\mathbb{R})$ algebra, generated by $L_0, L_{\pm 1}$ and $\bar L_0, \bar L_{\pm 1}$. The compactness of  $u(1)$ Cartan subalgebra of the $sl(2,\mathbb{R})$ algebras depends on the topology of $\Sigma$. For our case where we consider $\Sigma$ to be a compact surface, one of the two $u(1)$'s, the diagonal subalgebra of the Cartan of $sl(2,\mathbb{R})\oplus sl(2,\mathbb{R})$, is always compact. The other one is also compact if $\Sigma$ has toroidal topology.  The non-global part of the algebra generated by $L_{n}, \bar L_n,\ n\neq 0,\pm 1$, yields invertible transformations when $z=0$ and $z=\infty$ points are excluded.

By assuming that $Z$ and $W$ are state-independent, the adjusted bracket yields
\begin{equation}\label{KV-algebra-geometric}
    [\xi(  T_1, Z_1, W_1, Y_1^A), \xi( T_2, Z_2,  W_2, Y_2^A)]_{\text{\tiny Adj.B.}}=\xi(  T_{12}, Z_{12}, W_{12}, Y_{12}^A)\, ,
\end{equation}
where 
\begin{subequations}
\begin{align}
    &T_{12}=T_{1}\partial_{v}T_2-(1 \leftrightarrow 2)\, , \label{T12}\\
    &Z_{12}= T_1\partial_{v} Z_{2}-Z_{2}\partial_{v}T_{1}+Y_{1}^{A}\partial_{A}Z_{2}+Z_2 W_1 -(1 \leftrightarrow 2)  \, ,\\
    &W_{12}=T_1\partial_{v}W_2+Y_{1}^{A}\partial_{A}W_{2}-(1 \leftrightarrow 2)\, ,\\
     &Y_{12}^{A}={T_1\, \partial_v Y_{2}^{A}}+  Y_{1}^{B}\partial_{B}Y_{2}^{A}-(1 \leftrightarrow 2)\, .
\end{align}
\end{subequations}
\paragraph{Field transformations.} Under symmetry generating diffeomorphisms $\xi$, the fields transform as
\begin{subequations}
    \begin{align}
        &\delta_{\xi}\eta=T\, \partial_{v}\eta+2\, \eta \, \partial_{v}T - \eta \, W+\mathcal{L}_{Y}\eta \, , \\
        &\delta_{\xi}\Omega=T\, \partial_{v}\Omega+2\, \eta\, \lambda\, \sqrt{\Omega}\,  Z+\mathcal{L}_{Y}\Omega+D_{A}Y^A\, \Omega \, ,  \\
        &\delta_{\xi}\lambda=T \partial_{v}\lambda-\lambda \partial_{v}T+\mathcal{L}_{Y}\lambda+\frac{1}{2}\lambda\, D_{A}Y^A\, , \label{delta-lambda}\\
        &\delta_{\xi}\mathcal{U}^{A}=\mathcal{D}_{v}({Y}^A+T\mathcal{U}^A)\, , \label{delta-U}\\
        & \delta_{\xi}m=T \, \partial_{v}m+3\, m\, \partial_{v}T\, , \label{delta-xi-m} \\
        & \delta_{\xi} {\mathrm{P}}= T \partial_v {\mathrm{P}}+\mathcal{L}_{{Y}} {\mathrm{P}}- W- 2 \, Z \, e^{\Pi/2}\, ,\\
        &\delta_{\xi}q_{AB}=0\label{delta-xi-qAB}\, .
    \end{align}
\end{subequations}
In the above $Y^A$ is subject to \eqref{CKE-Y}. Recalling \eqref{mR3-EoM-2} and \eqref{beta-v}, the above imply
\begin{equation}
\delta_\xi \mathcal{M}_\circ =0\, , \qquad \delta_\xi \mathcal{R} = T \partial_v \mathcal{R} -\mathcal{R} \partial_v T \, ,
  \qquad  \delta_\xi \beta = T/\mathcal{R}\, .
\end{equation}

\subsection{Surface charges} 

Surface charge variations associated with the symmetry generator $\xi$ are given by \cite{Iyer:1994ys}
\begin{equation}
    \slashed{\delta} Q(\xi)=\bO^{\text{\tiny on-shell}}_{\text{\tiny reg.}}[g; \delta_\xi g,\delta g]\, .
\end{equation}
Recalling the transformation laws,
\begin{subequations}
    \begin{align}
         \delta_\xi \beta = &\ \tilde{T}:= T/\mathcal{R}\, , \\
        \delta_\xi{\mathrm{P}} = & -\tilde{W} \, , \\
      \delta_\xi \hat{\Omega}  =&\ \tilde{Z}\, ,
    \end{align}
\end{subequations}
one obtains
\begin{equation}
   \slashed{\delta} Q_{\text{\tiny {reg.}}}(\xi)=    \tilde{T} \, \delta \mathcal{M}_\circ +\frac{1}{16\pi G} \oint \d z \d{\bar{z}} \, \left(\tilde{Z} \, \delta \hat{\mathrm{P}}+\tilde{W}\,  \delta \hat{\Omega} \right) \, .
\end{equation}
In the charge expression above $\beta$ and its variation do not appear. That is, $\beta(v \neq v_b)$ is a pure gauge in solution phase space and can be fixed by a proper diffeomorphism. 

\paragraph{Integrable surface charge.}
In the slicing $\delta \tilde{W}=\delta \tilde{Z}=\delta \tilde{T}=0$, charge variations are integrable with the surface charges,
\begin{equation}\label{integrable-charge}
    Q_{\text{\tiny {reg.}}}(\xi)= \tilde{T}\,  \mathcal{M}_\circ +\frac{1}{16 \pi G}\oint \d z \d{\bar{z}}  \, \left(\tilde{Z} \, \mathrm{P}+\tilde{W}\,  \hat{\Omega} \right) \, .
\end{equation}
Furthermore, the adjusted Lie bracket of symmetry generators in the hatted slicing vanishes and hence
\begin{equation}
    \delta_{\xi_1} Q(\xi_2):=\{Q(\xi_1),Q(\xi_2)\}=K[\xi_2;\xi_1]\, ,\qquad \forall \, \xi_1,\xi_2\, ,
\end{equation}
where $K[\xi_2;\xi_1]$ is the Heisenberg central extension term 
\begin{equation}
    K[\xi_1;\xi_2]=-\frac{1}{16 \pi G} \oint \d z \d{\bar{z}} \,(\tilde{W}_1\, \tilde{Z}_2-\tilde{W}_2\, \tilde{Z}_1)\, .
\end{equation}
Explicitly, we have the charge algebra, 
\begin{equation}\label{Omega-P-algbera}
\begin{split}
    \{\hat{\Omega}(v;z,\bar z), \mathrm{P}(v;w,\bar w)\}= {16\pi G}\ \delta^2(z-w)\, ,\\
    \{\mathcal{M}_\circ, \hat{\Omega}(v;z,\bar z)\}=0= \{\mathcal{M}_\circ, \mathrm{P}(v;z,\bar z)\}\, .
\end{split}
\end{equation}
The above is clearly compatible with the Poisson brackets one can read from the on-shell symplectic form \eqref{symplectic-form-com}. 
\paragraph{Charges associated with superrotation symmetry generators.}
Now we aim to show that superrotations are a subsector of the solution phase space discussed above. To do so, we rewrite transformation laws as
\begin{subequations}\label{check-slicing}
    \begin{align}
        & \delta_{\zeta} {\mathrm{P}}= -\check{W}+\mathcal{L}_{\check{Y}} {\mathrm{P}} \, \qquad \check{W}= W+ 2 \, Z \, e^{\Pi/2}-T \mathcal{D}_v {\mathrm{P}}\, ,\\
        & \delta_{\zeta} \hat{\Omega}= \check{Z}+\mathcal{L}_{\check{Y}}\hat{\Omega} \, , \qquad \check{Z}=T\, \mathcal{D}_{v}\hat{\Omega}+2\, \eta\, \hat{\lambda}\, \hat{\Omega}\,  Z\, ,
    \end{align}
\end{subequations}
where we define $\check{Y}^A:= Y^A+T \mathcal{U}^A$. Then the surface charge variation can be written as
\begin{equation}\label{charge}
  \slashed{\delta}  Q_{\text{\tiny {reg.}}}(\xi)= \tilde{T}\,  \mathcal{M}_\circ +\frac{1}{16 \pi G}\oint \d z \d{\bar{z}}  \, \left(\check{Z} \, \delta {\mathrm{P}}+\check{W}\,\delta\hat{\Omega}- \check{Y}^A \delta (\hat{\Omega} \,\partial_A \mathrm{P})\right) \, .
\end{equation}
The above shows that for the special choice of symmetry generators $\tilde{T}=\check{W}=\check{Z}=0$, the surface charge expression yields superrotation charges
\begin{equation}\label{SR-charge}
    Q_{\text{\tiny {reg.}}}(Y)= -\frac{1}{16 \pi G}\oint \d z \d{\bar{z}}  \, {Y}^A  \, \hat{\Omega} \,\partial_A \mathrm{P} \, .
\end{equation}
One can simply check that the superrotation charges \eqref{SR-charge} yields the following algebra
\begin{equation}\label{SRC}
    \{Q_{\text{\tiny {reg.}}}(Y_1),Q_{\text{\tiny {reg.}}}(Y_2)\}=Q_{\text{\tiny {reg.}}}(Y_{12}^A)\, , \qquad Y_{12}^{A}=Y_{1}^{B}\partial_{B}Y_{2}^A-Y_{2}^{B}\partial_{B}Y_{1}^A\, ,
\end{equation}
without any central extension term. The above is the algebra one deduces for ${\cal J}_A:=\hat{\Omega} \,\partial_A \mathrm{P}$ using \eqref{Omega-P-algbera}. 

Having the boundary charges, we now discuss which of the parameters and functions in the solution space are physical and which can be removed by proper diffeomorphisms. The solution space is spanned by 2 numbers $\mathrm{g}$ and $\mathcal{M}_\circ$, a co-dimension 3 function $\beta(v)$ (or ${\cal R}(v)$), one vector field $\mathcal{U}^{A}(v,x^B)$, defined on transverse surface, and three co-dimension {1} functions  $\lambda(v,x^A),\Omega(v,x^A),\eta(v,x^A)$. The $\mathcal{U}^{A}(v,x^B)$ and $\lambda(v,x^A)$ are respectively subject to the CKE and the RT equation. 
\begin{itemize}
    \item 
$\beta$ (at an arbitrary advanced time $v_b$) is canonical conjugate to the ADM mass $\mathcal{M}_\circ$.
\item The two modes $\Omega(v,x^A),\eta(v,x^A)$ specify/label the boundary degrees of freedom and are physical.
\item $\lambda$ does not appear in the on-shell symplectic form but specifies the shape of the RT wavefront. 
\item $\mathcal{U}^{A}(v,x^B)$ does not explicitly appear in the surface charge expression, indicating that they are partially pure gauges.  We can choose $Y^A(v,x^A)$ such that ${\cal D}_v(Y^A(v,x^A)+T(v) {\cal U}^A(v,x^A))=0$. With this choice, \eqref{delta-U} implies $\delta_\xi{\cal U}^A=0,$ and hence ${\cal U}^A$ is fixed to a  $v$ independent vector on $\Sigma$ which satisfies CKE, i.e. ${\cal U}^A\partial_A={\cal U}(z)\partial_z+{\bar{\cal U}}(\bar z)\partial_{\bar z}$. Similarly, $Y^A$ is also reduced to  $Y^A \partial_{A}=Y^{z}(z)\partial_{z}+Y^{\bar{z}}(\bar{z})\partial_{\bar{z}}$. 
\item ``Superrotation generators'' $Y^z, Y^{\bar{z}}$ in the Heisenberg slicing do not yield any {new} independent charge. That is, in the check-slicing \eqref{check-slicing}, they lead to ``orbital (super)rotation''  \cite{Adami:2021nnf, Adami:2020ugu} charges \eqref{SRC} which form two copies of Witt algebra \eqref{SRC}. We note that the orbital superrotation charges are given in terms of the Heisenberg charges and are not functionally independent. See, however, the comment below. 
\item ``Superrotation generators'' $Y^z, Y^{\bar{z}}$ generators are not generically globally defined on $\Sigma$ so one may exclude them on the condition of smoothness of $\Sigma$. Therefore, assuming smoothness of $\Sigma$ and restricting to symmetry generators that respect this smoothness, we  have only 6 charges associated with the global $sl(2,\mathbb{R})\oplus sl(2,\mathbb{R})$ part of the superrotations. We note again that these charges in the slicing \eqref{check-slicing} are made from $\Omega, \mathrm{P}$ as in \eqref{SRC}. These 6 charges form $so(3,1)\simeq sl(2,\mathbb{R})\oplus sl(2,\mathbb{R})$ algebra. 

\end{itemize}

\section{Boundary energy-momentum tensor and fluid description, AdS case}\label{sec:BEMT}
In this section, we rewrite the dynamical equation \eqref{RT-eq} as a conservation of an energy-momentum tensor (EMT) at the asymptotic boundary. To do so, we first construct some geometric quantities at the asymptotic boundary. In this section, we restrict ourselves to the asymptotically AdS spacetimes and postpone the study of asymptotically flat cases to the next section.

\subsection{Geometric structure} 
We start with the asymptotic boundary metric for the AdS case $\Lambda=-3\ell^{-2}$, 
\begin{equation}
    \d{} \hat{\gamma}^2=\hat{\gamma}_{ab}\d x^a \d x^b=\Phi^2\left[\frac{1}{\ell^2 \lambda^2}\d v^2+q_{AB}(\d x^A+\mathcal{U}^A \d{}v) (\d x^B+\mathcal{U}^B \d{}v)\right]+\mathcal{O}(\Phi)\, ,
\end{equation}
where $x^{a}=\{v,x^A\}$.
To absorb the divergent coefficient in the induced metric, we define the conformal boundary as follows
\begin{equation}
    \gamma_{ab}=\lim_{r\to\infty}\frac{\varpi^2}{\Phi^2}\, \hat{\gamma}_{ab}\, ,
\end{equation}
where $\varpi=\varpi[\Omega,\lambda, \eta]$ is a positive definite generic functional on the solution phase space. Thus the boundary line element is
\begin{equation}\label{induced-metric-infty-ads}
    \d{} \gamma^2:=\gamma_{ab}\d x^a \d x^b=-\left(\frac{\varpi}{\ell \lambda}\right)^2 \d v^2+\varpi^2 \, q_{AB}(\d x^A+\mathcal{U}^A \d{}v) (\d x^B+\mathcal{U}^B \d{}v)\, .
\end{equation}
The determinant of the induced metric is $\sqrt{-\gamma}=\frac{\varpi^3\sqrt{q}}{\ell\lambda}$ and timelike vector field on the boundary is given by
\begin{equation}
    t_{a}\d{}x^{a}=-\frac{\varpi}{\ell\lambda}\d{}v\, , \qquad t^{a}\partial_{a}=\frac{\ell \lambda}{\varpi}(\partial_{v}-\mathcal{U}^A \partial_{A})\, .
\end{equation}
Furthermore, the induced metric on codimension 2 transverse surface is given by
\begin{equation}\label{induce-met-trans}
    \mathtt{h}_{ab}=\gamma_{ab}+t_{a}t_{b}\, , \qquad \d{}\mathtt{h}^2:=\mathtt{h}_{ab}\d{}x^a\d{}x^b=\varpi^2 \, q_{AB}(\d x^A+\mathcal{U}^A \d{}v) (\d x^B+\mathcal{U}^B \d{}v)\, .
\end{equation}
The covariant derivative of the timelike vector field can be decomposed as
\begin{equation}
   \check{\nabla}_a t_{b}=-t_{a} \, H_{b}+\frac{1}{2}\, \theta_t\,  \mathtt{h}_{ab}\, ,
\end{equation}
where $ \check{\nabla}_a$ is the covariant derivative with respect to $\gamma_{ab}$ and vector field $H_a$ and expansion $\theta_t$ are given as follows
\begin{subequations}
    \begin{align}
        & H_a \d x^a := \mathtt{h}_{ab} \, t^{c} \,\check{\nabla}_c t^{b} \d x^a =H_A \left(\d x^A + \mathcal{U}^A \d v\right)\, ,\\
        & H_{A} =-D_A \ln{\left|\frac{\lambda}{\varpi}\right|} \, , \qquad \theta_t= \frac{\ell \,\lambda}{\sqrt{q}\varpi^3}  \, \mathcal{D}_v \left(\sqrt{q}\varpi^2 \right)\, .
    \end{align}
\end{subequations}
So far the conformal factor $\varpi$ is unspecified. We may conveniently choose a conformal gauge such that $H_a=0$, or equivalently $\varpi=\lambda/\mathcal{R}$, leading to the following induced metric on the boundary
\begin{equation}\label{metric-01}
    \d{} \gamma^2=-\frac{1}{(\ell \mathcal{R})^2} \, \d v^2+\left(\frac{\lambda}{ \mathcal{R}}\right)^{2} q_{AB}(\d x^A+\mathcal{U}^A \d{}v) (\d x^B+\mathcal{U}^B \d{}v)\, ,
\end{equation}
and an identically vanishing $H_a$. Eq.~\eqref{beta-v} defines the \emph{natural time} coordinate then the boundary metric takes the form
\begin{equation}\label{gauge-fixed-metric}
    \d{} \gamma^2=-\ell^{-2} \d \beta^2+2 \tilde{\lambda}^2 \, (\d z+\tilde{\mathcal{U}}^z \d \beta) (\d {\bar{z}}+\tilde{\mathcal{U}}^{\bar{z}} \d \beta )\, ,
\end{equation}
where as defined before $\tilde{\lambda}={q}^{1/4}\lambda/\mathcal{R}$ and $\tilde{\mathcal{U}}^A=\mathcal{R}\, {\mathcal{U}}^A$. In our analysis below, we restrict ourselves to the specific choice $\varpi=\lambda/\mathcal{R}$. 
\subsection{Conserved energy-momentum tensor}\label{}
Motivated by surface charge analysis, let us define a traceless energy-momentum tensor as
\begin{equation}\label{EMT-1}
    \mathcal{T}^{ab}=-\frac{1}{2}\mathcal{M}_\circ\left( 3\, t^a t^b+\gamma^{ab} \right)\, , \hspace{.6 cm} \gamma_{ab} \mathcal{T}^{ab}=0\,  .
\end{equation}
As we see  Brown-York stress tensor \eqref{EMT-1} only depends on ADM mass ${\cal M}_\circ$ in the solution space.  Therefore, it is expected that the contribution of 
$ \mathcal{T}^{ab}$ to the on-shell symplectic form only involves the thermodynamic part in the symplectic two-form \eqref{thermo-Omega}, explicitly, 
\begin{equation}\label{tensor-rep}
    \bO^{\mathrm{\tiny{thermo}}}_{\text{\tiny reg.}}=-\frac{\ell}{8\pi}\int \d{}^{3}x\, \delta\left(\sqrt{-\gamma}\mathcal{T}^{ab}\right) \wedge \delta\gamma_{ab}\, .
\end{equation}
The point is that $\mathcal{T}^{ab}$ is not conserved and its non-conservation may be compensated by a tensor, say $\mathcal{C}^{ab}$:
\begin{equation}\label{conservation-eq-1}
    \check{\nabla}_a \mathcal{T}^{ab}=\check{\nabla}_a\mathcal{C}^{ab}\, ,
\end{equation}
where the ``compensatory'' energy-momentum tensor is  the symmetric-traceless tensor
\begin{equation}
    \mathcal{C}^{ab}=\frac{\ell}{2 G}\left[C^{(ab)c}\, t_{c}+C^{cd(a}t^{b)}t_{c}t_{d}\right]\, ,
\end{equation}
Here $C_{abc}$ denotes the Cotton tensor, \footnote{Note that $\gamma^{ab} C_{abc}=0$, $C_{abc}=-C_{acb}$ and $ C_{(abc)}=0$. }
\begin{equation}
    C_{abc}=\check{\nabla}_c\mathrm{R}_{ab}-\check{\nabla}_b\mathrm{R}_{ac}+\frac{1}{4}\left(\gamma_{ac}\check{\nabla}_b\mathrm{R}-\gamma_{ab}\check{\nabla}_c\mathrm{R}\right)\, ,
\end{equation}
where $\mathrm{R}_{ab}$ and $\mathrm{R}$ are Ricci tensor and Ricci scalar of codimension-one conformal boundary $\gamma_{ab}$ respectively.  
Recalling that ${\mathcal{C}}_{ab}$ is traceless, ${\gamma}^{ab}{\mathcal{C}}_{ab}=0$, one can define a traceless divergent-free energy-momentum tensor,
\begin{equation}\label{total-EMT}
    \text{T}^{ab}:={\mathcal{T}}^{ab}-{\mathcal{C}}^{ab}\, , \hspace{1 cm} \check{\nabla}_a\text{T}^{ab}=0\, , \hspace{1 cm} \text{T}:={\gamma}^{ab}\text{T}_{ab}=0\, .
\end{equation}
We note that $\check{\nabla}_a\text{T}^{ab}=0$ is a consequence of the RT equation. \footnote{We note that conservation of the total boundary energy-momentum tensor is a direct consequence of the boundary symmetries \eqref{BSG-1}. In our case, the symmetry generators are only a subset of boundary diffeomorphisms. However, they are sufficient to guarantee the conservation of our specific boundary stress tensor.}
One can decompose the total energy-momentum tensor $\text{T}^{ab}$ as we do for a generic fluid, 
\begin{equation}\label{total-EMT-AdS-decom}
\text{T}^{ab}=\left(-\mathcal{E} + \mathcal{P}\right) \, t^a \, t^b+2\mathcal{J}^{(a}\, t^{b)}+\mathcal{P} \gamma^{ab}+\sigma^{ab}\, ,
\end{equation}
where energy density $\mathcal{E}$ and pressure $\mathcal{P}$ are
\begin{equation}
    \mathcal{E}= \mathcal{M}_{\circ}\, , \hspace{1 cm} \mathcal{P}=-\frac{1}{2} \mathcal{M}_{\circ}\, , \hspace{1 cm} \mathcal{E}+2\mathcal{P}=0\, .
\end{equation}
The last equation is the equation of state for a $3d$ conformal fluid. The momentum flow is also given by
\begin{equation}\label{RT-momentum-current}
\mathcal{J}^{a}\partial_a:= -\mathtt{h}^a{}_b t_c \text{T}^{bc} \partial_a = \mathcal{J}^{A}\partial_A \, , \qquad
    \mathcal{J}^{A}=\frac{\ell}{8 G\varpi^2}  D^{A}\left[\varpi^{-2}\left(D^2\ln{\varpi^2}-R[q]\right)\right]\, ,
\end{equation}
and the shear-stress tensor is as follows
\begin{equation}\label{RT-shear}
   \sigma^{ab}\partial_a \otimes \partial_b=(\mathtt{h}^{a}_{c}\mathtt{h}^{b}_{d}-\frac{1}{2}\mathtt{h}^{ab}\mathtt{h}_{cd})\text{T}^{cd}\partial_a \otimes \partial_b= \sigma^{AB}\partial_A \otimes \partial_B\, ,
\end{equation}
where
\begin{equation}
    \sigma^{AB}=\frac{\ell}{{2} G \varpi^5} \left( D^{\langle A}\varpi D^{B \rangle} \theta_t -\varpi\,  D^{\langle A} D^{B \rangle} \theta_t \right) \, .
\end{equation}
Here $\langle \cdot , \cdot \rangle $ denotes the symmetric-traceless part {(see below equation \eqref{NAB-l-leading-r} for its definition)}. Recalling \eqref{EMT-1}, one observes that the ${\cal J}^a, \sigma^{ab}$ terms in \eqref{total-EMT-AdS-decom} come from the ``compensatory'' term ${\cal C}^{ab}$. RT equation \eqref{RT-eq-2} for $\varpi=\lambda/{\cal R}$ may then be written as a conservation equation, 
\begin{equation}\label{CQ-01}
     t_a\check{\nabla}_b \text{T}^{ab}=0\, . 
\end{equation}

We close this part mentioning that the total energy-momentum tensor \eqref{total-EMT} which is derived from the analysis of surface charges \eqref{integrable-charge}, symplectic form \eqref{symplectic-form-com}, and  RT equation of motion \eqref{RT-eq}, coincides with the one in \cite{BernardideFreitas:2014eoi} where the usual AdS/CFT technique \cite{Balasubramanian:1999re} was used to obtain the boundary energy-momentum tensor.  Our result \eqref{total-EMT} generalizes analysis of \cite{BernardideFreitas:2014eoi} to $m=m(v)$ and $\mathcal{U}^A \neq 0$.

\paragraph{Symplectic form.} 
In terms of these energy-momentum tensors, the symplectic form is given by
\begin{equation}\label{symplectic-form-hydrodynamic}
    \begin{split}
        \bO_{_{\mathcal{C}}}&=\int_{\mathcal{C}_{r}}\d{}^3 x\left[-\frac{\ell}{8\pi}\delta(\sqrt{-{\gamma}}{\mathcal{T}}^{ab})\wedge\delta{\gamma}_{a b}+\frac{1}{16\pi G}{\partial_{v}(\delta\hat{\Omega}\wedge\delta\mathrm{P})}\right]\\
        &=\int_{\mathcal{C}_{r}}\d{}^3 x\left[-\frac{\ell}{8\pi}\delta(\sqrt{-{\gamma}}{\text{T}}^{ab})\wedge\delta{\gamma}_{a b}+\frac{1}{16\pi G}{\partial_{v}}{(\delta\hat{\Omega}\wedge\delta\mathrm{P})}\right]\, .
    \end{split}
\end{equation}
That is, the compensatory term ${\cal C}^{ab}$ and hence ${\cal J}^A, \sigma^{AB}$, which carry the information about the RT mode and shape of the RT wavefront, do not contribute to the on-shell symplectic form. The key takeaway from the analysis above is that although the momentum current \eqref{RT-momentum-current} and shear tensor \eqref{RT-shear} of RT gravitational waves contribute to the total boundary energy-momentum tensor \eqref{total-EMT}, they do not contribute to the symplectic form \eqref{symplectic-form-hydrodynamic}. In hydrodynamics terminology, it is pertinent to note that the symplectic form exclusively incorporates zero-order hydrodynamics variables, namely energy and pressure, along with their conjugates. Notably, higher-order information, such as the RT momentum current and RT shear, is not encompassed within the symplectic form. This aligns completely with the conclusions drawn in \cite{Adami:2023wbe} which indicates the RT mode appears as the kernel of the Carrollian structure on null boundary solution space.\footnote{Note that compared to \cite{Adami:2023wbe}, while we consider a generic causal (and not just null) boundary, we have turned off the usual tensor modes, denoted by $N_{AB}$ in \cite{Adami:2023wbe}. The RT mode (in our current analysis encoded in $\lambda$, couple to $N_{AB}$ modes and hence do appear in the bulk (codimension 1 integral part) of the symplectic form.} Analysis here reveals that this property holds even for generic boundaries. 
We also observe that the codimension 2 modes do not contribute to the total energy-momentum tensor. The same structure appears in $3d$ Einstein gravity \cite{Adami:2023fbm}.
\subsection{Conserved current}
We have demonstrated that the RT equation can be reformulated as the conservation of the energy-momentum tensor. This reformulation enables us to express the symplectic form of the RT solution space in a canonical form, as shown in \eqref{symplectic-form-hydrodynamic}. Furthermore, we highlight a novel approach to express the RT equation and its symplectic form using a conserved current. This elegant representation is made possible by the scalar nature of the RT radiative mode, a distinctive characteristic that distinguishes it from generic spacetimes with radiative tensor modes, where such formulation is unattainable. The unique scalar property of the RT radiative mode plays a pivotal role in facilitating the identification of an integrable slicing in the presence of the radiative mode.

It is straightforward to demonstrate that the thermodynamics part of the symplectic form, as expressed in \eqref{thermo-Omega}, can be rewritten as
\begin{equation}\label{vector-rep}
    \bO^{\mathrm{\tiny{thermo}}}_{\text{\tiny reg.}}=\frac{\ell}{4\pi}\int \d{}^{3}x\, \delta(\sqrt{-\gamma}\pi^a)\wedge \delta t_{a}\, ,
\end{equation}
where $\pi^a$ is a timelike current parallel to $t^a$,
\begin{equation}
    \pi^{a}\partial_{a}=\mathcal{M}_{\circ} t^{a}\partial_{a}\, .
\end{equation}
To discuss the conservation equation, let us rewrite \eqref{CQ-01} as
\begin{equation}\label{}
     \check{\nabla}_a (  \text{T}^{ab}t_b )=\text{T}^{ab}\check{\nabla}_a t_b =-\frac{1}{2}\mathcal{M}_\circ \theta_t\, . 
\end{equation}
Since $\theta_t= \check{\nabla}_a t^a$, the above yields a conserved current 
\begin{equation}
  \Pi^a := \pi^a - \frac{2}{3} \mathcal{J}^a \, , \qquad  \check{\nabla}_a \Pi^a = 0 \, ,
\end{equation}
where $\mathcal{J}^a$ is the RT momentum flow \eqref{RT-momentum-current}. The above shows that the RT equation can be written as a conservation equation for a conserved current. It is worth noting that the momentum flow $\mathcal{J}^a$ serves as \emph{compensating vector} and its role is to ensure the RT equation written as $\check{\nabla}_a \Pi^a = 0$. Finally, we rewrite the total symplectic form on the asymptotic AdS boundary as
\begin{equation}\label{SF-VR-AdS}
    \begin{split}
        \bO_{_{\mathcal{C}}}&=\int_{\mathcal{C}}\d{}^3 x\left[\frac{\ell}{4\pi} \delta(\sqrt{-\gamma}\pi^a)\wedge \delta t_{a}+\frac{1}{16\pi G}{\partial_{v}(\delta\hat{\Omega}\wedge\delta\mathrm{P})}\right]\\
        &=\int_{\mathcal{C}}\d{}^3 x\left[\frac{\ell}{4\pi} \delta(\sqrt{-\gamma}\Pi^a)\wedge \delta t_{a}+\frac{1}{16\pi G}{\partial_{v}}{(\delta\hat{\Omega}\wedge\delta\mathrm{P})}\right]\, .
    \end{split}
\end{equation}

In summary, there are two alternative representations of the thermodynamic component within the boundary symplectic 2-form for RT solutions \eqref{thermo-Omega}: the vector representation \eqref{vector-rep} and the tensor representation \eqref{tensor-rep}. Although they characterize the same boundary system, each offers distinct advantages. Both representations feature the ADM mass. The vector representation incorporates momentum flow, while the tensor representation encompasses the shear-stress tensor as well. 

\section{Energy-momentum tensor and hydrodynamic description, flat case}\label{T-flat}
In this section, we take similar steps as the previous section for asymptotically flat spacetimes. The main difference is that the boundary metric is a Carrollian geometry.

\subsection{Geometric structure}
In the asymptotically flat spacetimes with $\Lambda=0$, the line element at null infinity can be found by setting $\Lambda=0$ in $V$ function \eqref{V-expression}, while taking large $r$ limit. This yields \eqref{metric-asymptotic-flat-bdry} which in the leading order is, 
\begin{equation}\label{bdry-metric-null}
    \d{} \gamma^2=\gamma_{ab}\d x^a \d x^b=\varpi^2 \, q_{AB}(\d x^A+\mathcal{U}^A \d{}v) (\d x^B+\mathcal{U}^B \d{}v)\, .
\end{equation}
The same induced metric can be obtained by taking the limit $\ell \to \infty$ of \eqref{induced-metric-infty-ads}. It is worth mentioning that the cosmological constant does not appear in RT equation \eqref{RT-eq}. While the conformal boundary of the RT spacetime in the flat case $\Lambda=0$ is a $3d$ hypersurface, its induced metric is degenerate. It means there exists a kernel vector such that
\begin{equation}\label{l-kernel}
    l^{a}\partial_{a}=\alpha^{-1} \, (\partial_{v}-\mathcal{U}^A \partial_{A})\, , \qquad {\gamma}_{ab} l^a=0\, ,
\end{equation}
where $\alpha$ is an arbitrary function of $v,x^A$. 
The  degenerate metric \eqref{bdry-metric-null} together with the kernel vector \eqref{l-kernel} constitutes the \textit{weak} Carrollian structure (\emph{c.f.} \cite{Duval:2014uoa, Duval:2014uva, Duval:2014lpa, Henneaux:1979vn,Henneaux:2021yzg, Ciambelli:2019lap,deBoer:2021jej, deBoer:2017ing} for more discussions). To study the Carrollian structure, we also need to define the Ehresmann connection (or clock one-form)
\begin{equation}\label{Ehresmann}
    n_{a}\d{}x^a=-\alpha \d{}v+n_A (\d x^A +\mathcal{U}^A \d v) \, , \qquad l^{a}n_{a}=-1\, .
\end{equation}
The ternary quantities $\{\gamma_{ab},l^{a},n_{a}\}$ formed a \textit{ruled} Carrollian structure. 
The projection map 
\begin{equation}
P^{a}{}_{b}:=\delta_{b}^{a}+n_{b}l^{a}\, , \qquad P^{a}{}_{b}l^{b}=0\, , \qquad P^{a}{}_{b}n_{a}=0\, ,
\end{equation}
allows us to define a symmetric tensor with upper indices $h^{ab}$ such that $h^{ac}\gamma_{cb}=P^{a}{}_{b}$. Without loss of generality, we set $n_{A}=0$, and to specify $h^{ab}$ uniquely we may impose $h^{ab}n_an_b=0$. Thus components of $h^{ab}$ read as
\begin{equation}
   h^{vv}=0 \, , \qquad h^{vA}=0\, , \qquad h^{AB}= \varpi^{-2}\, q^{AB} \, .
\end{equation}
The minors of $\gamma_{ab}$ are defined as $\mathfrak{g}^{ab}=1/2 \varepsilon^{acd}\varepsilon^{bef}\gamma_{ce}\gamma_{df}$. Because $\mathfrak{g}^{ab}\gamma_{bc}=0$ then the minors can be written as $\mathfrak{g}^{ab}=\alpha^{2}\Breve{\gamma} \, l^a l^b$ where $\Breve{\gamma}= -\varpi^4\, q $ plays the role of the determinant of metric when the metric is degenerate.
\paragraph{Null boundary connection.}
We define the Carroll-Affine connection as
\begin{equation}\label{null-connection}
   \Breve{\Gamma}^c_{ab}:= \frac{1}{2}h^{cd}\left(\partial_{a}\gamma_{bd}+\partial_{b}\gamma_{ad}-\partial_{d}\gamma_{ab}\right)\, ,
\end{equation}
and hence Carroll covariant derivative for a generic tensor $\mathcal{X}^{a\cdots}_{b \cdots}$ can be defined as
\begin{equation}
    \Breve{\nabla}_c \mathcal{X}^{a\cdots}_{b \cdots}=\partial_c \mathcal{X}^{a\cdots}_{b \cdots} + \Breve{\Gamma}^a_{cd} \mathcal{X}^{a\cdots}_{b \cdots}+\cdots- \Breve{\Gamma}^d_{cb} \mathcal{X}^{a\cdots}_{d \cdots}-\cdots \, .
\end{equation}
While we use the prescription of \cite{Bergshoeff:2017btm} to compute the Riemann tensor, the connection defined above differs. The connection we defined here reduces to spatial and temporal connections defined in \cite{Bagchi:2023rwd} when we set $\varpi=1$ and $\mathcal{U}^A=0$. An important point is that connection \eqref{null-connection} is not compatible with degenerate metric $\gamma_{ab}$, i.e. we have
\begin{equation}
    \Breve{\nabla}_c \gamma_{ab}=  \theta_l \, \gamma_{c(a} n_{b)}\, ,
\end{equation}
where $\theta_l$ is expansion of kernel vector $l^a$, that is
\begin{equation}
    \theta_l:= P^a{}_{b} \Breve{\nabla}_a l^b=2\, \frac{\mathcal{D}_v (\varpi \sqrt{q})}{\alpha \varpi\sqrt{q}} \, .
\end{equation}
In the context of the discussion in section \ref{sec:Null energy-momentum tensor}, it is pertinent to note that while there exists a conformal frame $\varpi=1$ in which the metric and connection are compatible, to express the RT equation as a conservation law for a specified energy-momentum tensor, and to express the symplectic form in terms of the given tensor, we shall fix conformal factor $\varpi$ in particular way. We also note that $\Breve{\nabla}_c h^{ab}=0$.

\paragraph{Null boundary curvature.}
Given the Carroll-Affine connection \eqref{null-connection}, we can define the corresponding null Carroll-Riemann tensor
\begin{equation}
    \Breve{R}^{d}{}_{abc}:=\partial_{b}\Breve{\Gamma}^{d}{}_{ac}-\partial_{c}\Breve{\Gamma}^{d}{}_{ab}+\Breve{\Gamma}^{e}_{ac}\Breve{\Gamma}^{d}_{be}-\Breve{\Gamma}^{e}_{ab}\Breve{\Gamma}^{d}_{ce}\, .
\end{equation}
Consequently, the Carroll-Ricci tensor can be defined as $\Breve{R}_{ab}:= \Breve{R}^{c}{}_{acb}$. To define the Ricci scalar we need to contract the Ricci tensor indices with the inverse metric. On the null infinity in the conformal gauge, the metric is degenerate and hence non-invertable.  Instead, we define the Carroll-Ricci scalar as $\Breve{R}:=h^{ab}\Breve{R}_{ab}$.
\paragraph{Carroll-Cotton tensor.}
For later applications, we also define the Carroll-Cotton tensor
\begin{equation}
    \Breve{C}_{abc}=\Breve{\nabla}_c \Breve{R}_{ab}-\Breve{\nabla}_b\Breve{R}_{ac}+\frac{1}{4}\left(\gamma_{ac}\Breve{\nabla}_b\Breve{R}-\gamma_{ab}\Breve{\nabla}_c\Breve{R}\right)\, .
\end{equation}
Once we construct the null boundary Brown-York energy-momentum tensor we will observe that it involves Carroll-Cotton tensor.
\paragraph{Null boundary freedoms.}
The geometric structure of the null boundary discussed above emerges with two freedoms: the conformal factor $\varpi$ in \eqref{bdry-metric-null} and the boost parameter $\alpha$ in the definitions of the kernel \eqref{l-kernel} (and subsequently in the Ehresmann connection \eqref{Ehresmann}). In the following, we will appropriately fix these parameters.

\subsection{Null energy-momentum tensor}\label{sec:Null energy-momentum tensor} 
We now construct the null energy-momentum tensor. To do so, we need first to fix the boost freedom and the conformal factor as $\alpha=1$ and $\varpi=\lambda$ respectively. Motivated by the charge analysis \eqref{charge}, and recalling the relation between $m$ and $\mathcal{M}_{\circ}$, namely \eqref{mR3-EoM-2}, we define a trace-free null energy-momentum tensor as
\begin{equation}
    \Breve{\mathcal{T}}^{a}{}_{b}=-\frac{1}{2}\, m \left( P^{a}{}_{b}+2 l^{a}n_{b}\right)\, , \qquad \Breve{\mathcal{T}}:=\Breve{\mathcal{T}}^{a}{}_{a}=0\, .
\end{equation}
The (2,0)-tensor form of the null energy-momentum tensor is
\begin{equation}\label{EMT-null-1}
    \Breve{\mathcal{T}}^{ab}:=h^{bc}\Breve{\mathcal{T}}^{a}{}_{c}={-\frac{1}{2}}m\, h^{ab}\, ,
\end{equation}
where $h^{ab}n_{b}=0$ was used. 
The above version of the null energy-momentum tensor allows us to write the thermodynamic symplectic form in the canonical form
\begin{equation}\label{SF-can-null-1}
    \bO^{\mathrm{\tiny{thermo}}}_{\text{\tiny reg.}}={-\frac{1}{8\pi}}\int \d{}^{3}x\, \delta\left(\sqrt{-\Breve{\gamma}} \, \Breve{\mathcal{T}}^{ab}\right) \wedge \delta \gamma_{ab}\, ,
\end{equation}
while the $(1,1)$-tensor version of the null energy-momentum tensor is useful to write down the conservation equation. One would infer that $\Breve{\nabla}_b \Breve{\mathcal{T}}^{b}{}_{a}=\Breve{\nabla}_b \Breve{\mathcal{C}}^{b}{}_{a}$ where the compensatory tensor $\Breve{\mathcal{C}}^{a}{}_{b}$ is given by
\begin{equation}
     \Breve{\mathcal{C}}^{a}{}_{b}={-\frac{1}{{2}G}}\, n_{b}h^{ac}h^{de}\Breve{C}_{dce}\, ,
\end{equation}
yields RT equation \eqref{RT-eq}. Therefore, we can define a trace-free and divergence-free null energy-momentum tensor as
\begin{equation}\label{total-null-EMT-1}
    \Breve{\text{T}}^{a}{}_{b}:=\Breve{\mathcal{T}}^{a}{}_{b}- \Breve{\mathcal{C}}^{a}{}_{b}\, , \qquad  \Breve{\nabla}_a \Breve{\text{T}}^{a}{}_{b}=0\, , \qquad \Breve{\text{T}}:=\Breve{\text{T}}^{a}{}_{a}=0\, .
\end{equation}
The total null energy-momentum tensor defined above can be decomposed as
\begin{equation}\label{NEMT-D}
    \Breve{\text{T}}^{a}{}_{b}=-\Breve{\mathcal{E}} \, l^a \, n_b+\Breve{\mathcal{J}}^{a}\, n_{b}+\Breve{\mathcal{P}} \, P^{a}{}_{b}\, ,
\end{equation}
where
\begin{equation}\label{ES}
    \Breve{\mathcal{E}}=m\, , \qquad  \Breve{\mathcal{P}}=-\frac{m}{2}\, , \qquad \Breve{\mathcal{E}}+2\Breve{\mathcal{P}}=0\, ,
\end{equation}
and 
\begin{equation}
    \Breve{\mathcal{J}}^a:=\Breve{\mathcal{C}}^{a}{}_{b}\, l^{b} \, , \qquad  \Breve{\mathcal{J}}^a \partial_{a}=\frac{1}{8G\lambda^2}D^A\left[\lambda^{-2}\left(D^2\ln\lambda^2-R[q]\right)\right]\partial_{A}\, .
\end{equation}
Eq.~\eqref{NEMT-D} implies that the boundary fluid is a conformal Carrollian one and the equation of state is given by \eqref{ES}, \emph{cf.} \cite{Ciambelli:2017wou}. {We note that the information in the RT mode $\lambda$ is in the compensatory piece $\Breve{\mathcal{C}}^{a}{}_{b}$ which yields the $\Breve{\mathcal{J}}^a$ and its role is to reproduce the RT equation from $\Breve{\nabla}_a \Breve{\text{T}}^{a}{}_{b}=0$.}

It is important to note that unlike the AdS case described in \eqref{total-EMT-AdS-decom}, the total null energy-momentum tensor doesn't include the tensor mode. Essentially, this means that the RT-shear is missing from the null energy-momentum tensor. This feature is consistent with the observation in the energy-momentum tensor on thin null shells, as discussed by Barrab\'es and Israel \cite{PhysRevD.43.1129}.

We wrap up this section by expressing the symplectic form in terms of the null energy-momentum tensors  
\begin{equation}\label{}
    \begin{split}
        \bO_{_{\mathcal{N}}}&=\int_{\mathcal{N}}\d{}^3 x\left[{-\frac{1}{8\pi}}\,\delta\left(\sqrt{-\Breve{\gamma}}\, \Breve{\mathcal{T}}^{ab}\right) \wedge \delta \gamma_{ab}+\frac{1}{16\pi G}{\partial_{v}(\delta\hat{\Omega}\wedge\delta\mathrm{P})}\right]\\
        &=\int_{\mathcal{N}}\d{}^3 x\left[{-\frac{1}{8\pi}}\delta(\sqrt{-\Breve{\gamma}}\, \Breve{\text{T}}^{ab})\wedge\delta{\gamma}_{a b}+\frac{1}{16\pi G}{\partial_{v}}{(\delta\hat{\Omega}\wedge\delta\mathrm{P})}\right]\, ,
    \end{split}
\end{equation}
where we defined $\Breve{\text{T}}^{ab}:=h^{bc}\Breve{\text{T}}^{a}{}_{c}$. As we see the compensatory part $\Breve{\mathcal{C}}^{a}{}_{b}$ and hence $\Breve{\mathcal{J}}^a$ do not appear in the on-shell symplectic form.
\subsection{Null boundary current} 
In this section, we construct a null boundary current with two key criteria: \emph{i}) ensuring that its conservation leads to  RT equation \eqref{RT-eq}, and \emph{ii}) expressing the thermodynamic component of the symplectic form \eqref{symp-form-reg} in terms of the conserved current. To do this, we fix the boost freedom as $\alpha=\mathcal{R}^{-1}$. The desired current is
\begin{equation}\label{null-current}
    \Breve{\pi}^{a}\partial_{a}=\mathcal{M}_{\circ}\mathcal{R}^{-1} \left(\partial_{v}- \mathcal{U}^A \partial_A \right)\, .
\end{equation}
The above current is proportional to the kernel vector field \eqref{l-kernel}. Now we can express the thermodynamic symplectic form \eqref{symp-form-reg} in terms of the current \eqref{null-current} as
\begin{equation}
    \bO^{\mathrm{\tiny{thermo}}}_{\text{\tiny reg.}}=\frac{1}{4\pi}\int \d{}^{3}x\, \delta(\sqrt{-\Breve{\gamma}}\,\Breve{\pi}^a)\wedge \delta n_{a}\, .
\end{equation}
The conformal factor here is chosen as $\varpi=\lambda\mathcal{R}^{-1/2}$.

Considering the Carroll-covariant derivative of the given current we can find that $\Breve{\nabla}_{a}\Breve{\pi}^a=\Breve{\nabla}_{a}\Breve{\mathcal{B}}^a$, where 
\begin{equation}
    \Breve{\mathcal{B}}^{a}=\frac{1}{3G}h^{ab}h^{cd}\Breve{C}_{cbd}\, ,
\end{equation}
stands for compensating vector field for the null boundary.
Thus total divergence-free conserved current can be defined as
\begin{equation}
    \Breve{\Pi}^{a}:= \Breve{\pi}^{a}-\Breve{\mathcal{B}}^{a}\, , \qquad \Breve{\nabla}_{a}\Breve{\Pi}^{a}=0\, .
\end{equation}
The equation above is nothing but the RT equation \eqref{RT-eq}, as we expected.
The symplectic form on the asymptotic null boundary takes the following form
\begin{equation}\label{symplectic-form-null}
    \begin{split}
        \bO_{_{\mathcal{N}}}&=\int_{\mathcal{N}}\d{}^3 x\left[\frac{1}{4\pi} \delta(\sqrt{-\Breve{\gamma}}\,\Breve{\pi}^a)\wedge \delta n_{a}+\frac{1}{16\pi G}{\partial_{v}(\delta\hat{\Omega}\wedge\delta\mathrm{P})}\right]\\
        &=\int_{\mathcal{N}}\d{}^3 x\left[\frac{1}{4\pi} \delta(\sqrt{-\Breve{\gamma}}\,\Breve{\Pi}^a)\wedge \delta n_{a}+\frac{1}{16\pi G}{\partial_{v}}{(\delta\hat{\Omega}\wedge\delta\mathrm{P})}\right]\, .
    \end{split}
\end{equation}

\subsection{Comparison between vector and tensor representations} 

As we have observed, the determination of the boost parameter and conformal factor differs between the vector and tensor representations:
\begin{itemize}
    \item \textit{Vector representation.} Expressing the RT equation in terms of a conserved current, and formulating the symplectic form accordingly, uniquely determine the conformal factor as $\varpi=\lambda\mathcal{R}^{-1/2}$ and the boost parameter as $\alpha=\mathcal{R}^{-1}$.
    \item \textit{Tensor representation.} By presenting the thermodynamic symplectic form in the canonical form \eqref{SF-can-null-1} and ensuring that the divergence of the null energy-momentum tensor \eqref{EMT-null-1} yields the RT equation, we uniquely determine the conformal factor as $\varpi=\lambda$ and the boost parameter as $\alpha=1$.
\end{itemize}

In the null case, there is a clear distinction from the AdS case: we opt for a consistent choice of the conformal factor for both the conserved current and the conserved energy-momentum tensor. Moreover, it's worth noting that the vector representation can be derived from the tensor representation, as discussed previously.


\section{Discussion}\label{sec:discusion}

We have constructed the most general family of solutions in vacuum Einstein-$\Lambda$ theory which admit a twist and shear-free null geodesic congruence with non-zero expansion. That is, we have extended the Robinson-Trautman (RT) solution to incorporate not only the ADM mass and RT scalar mode $\lambda (v,z,\bar z)$ (which parameterizes the time-dependent shape of the compact wavefront) but also two boundary modes and a vector field indicative of the local angular velocity of causal hypersurfaces. The analysis has revealed that the vector field, defined on a codimension one surface, satisfies the conformal Killing equations on the wavefront. Within this class of solutions, the wavefront, defined as a smooth compact $2d$ surface, exhibits versatility, allowing for an arbitrary genus $\mathrm{g}$. Additionally, the cosmological constant $\Lambda$ can vary, being positive, zero, or negative.

We have identified a collection of symmetries, encompassing a codimension 3 time-supertranslation, codimension 1 radial-supertranslation, and Weyl scaling, along with codimension 1 superrotations that are subjected to the conformal Killing equation.

Our investigation into surface charges has shown that only specific combinations, dependent on the chosen slicing, result in non-zero charges for the given boundary symmetries. 
A pivotal aspect of our study is the unveiling of a Heisenberg slicing, in which the surface charge algebra adopts the form of a Heisenberg algebra. Within this framework, the ADM mass emerges as a Casimir of the algebra.

Moreover, we have thoroughly explored the inclusion of conventional superrotations within the spectrum of surface charges. We have discussed that assuming smoothness of the wavefront $\Sigma$ only there are 6 charges forming $so(3,1)$ algebra in this sector and moreover, these 6 charges are not independent of the Heisenberg boundary charges. 

We delved into the generalized RT solution within the framework of the fluid/gravity correspondence. Through a careful analysis of the symplectic form and surface charge content, we have developed its corresponding fluid dual for both asymptotically AdS and flat spacetimes. This result paves the way for future studies of holography in these settings.

Using the formalism and results of our previous papers \cite{Adami:2023fbm, Adami:2023wbe}, we have decomposed the on-shell symplectic form over our solution space into three distinct components: the bulk (codimension 1) and the boundary (codimension 2) and thermodynamic (codimension 4) pieces. The latter involves the ADM mass and its conjugate, which may be viewed as the inverse of a temperature and the boundary symplectic form captures two Heisenberg soft hair modes of the solution space. 

Moreover, we have shown that the thermodynamic symplectic form can be expressed through a conserved energy-momentum tensor and/or conserved current, applicable to both asymptotically AdS and flat spacetimes. This discovery carries profound implications for our comprehension of flat space holography.

The RT mode, the function $\lambda (v,z, \bar z)$, is fully specified by the shape of the wavefront at any given time $v_b$, through $\lambda (v_b,z, \bar z)$, via the RT equation. Nonetheless, the RT mode does not appear in the on-shell symplectic form. This is in accord with the Carrollian nature of the solution phase space discussed in \cite{Adami:2023wbe} where the RT mode was identified as the kernel vector. This fact has two important physical consequences: (\emph{i}) One can't have simple memory effects, as we have for usual tensorial GW modes, for the RT mode. As the RT GW mode does not couple to the soft hair, in the absence of the symmetric-traceless tensor modes. (\emph{ii}) There is no symplectic flux associated with the RT mode. Therefore, the presence or absence of RT mode does not alter the integrability of surface charges. It has been discussed that the condition for integrability of boundary/surface charges is the absence of bulk modes \cite{Grumiller:2020vvv, Adami:2020ugu, Adami:2021sko, Adami:2021nnf, Adami:2021kvx, Taghiloo:2022hxc}. Our result here loosens this condition: there could be bulk modes (like the RT mode) that do not couple to the boundary modes. Given the two points above, one can arrive at a new statement for the integrability condition and existence of the memory effect:
\begin{center}
    \textit{`` There exists appropriate solution space slicing in which boundary/surface charges are integrable, if the bulk modes do not contribute to the symplectic flux. Such  bulk modes do not lead to memory effects.''  }
\end{center}

In our analysis here we focused on the RT modes with smooth wavefront (while allowing for arbitrary genus). It follows from our analysis that the genus of the wavefront can't change as the RT wave propagates in spacetime. That is, the genus of the wavefront is a part of the initial (Cauchy) data that defines the RT wave. One may extend our class of solutions by allowing the wavefront to have punctures. This case happens when we have cosmic strings that can pierce through the wavefront. Such solutions may be found in the class of C-metrics e.g. see  \cite{hoenselaers2005axisymmetric}.  For such cases, the initial and final metrics in the evolution process of cosmic strings, via black hole pair nucleation, differ by a finite superrotation. It is interesting to extend our analysis here to include cosmic strings attached to the wavefront. In this case, the total divergence terms in the symplectic form may not be dropped and they lead to new codimension 2 terms in the symplectic form and hence surface charges. In particular, these new terms would involve the RT mode $\lambda$, opening the way to ``probe'' the RT mode using boundary modes (in particular the superrotation modes). The first steps in this direction were briefly discussed in the appendix of \cite{Strominger:2016wns}.

\section*{Acknowledgement}
We would like to thank  Glenn Barnich, Piotr T. Chru\'sciel, Luca Ciambelli, Marc Geiller, C\'eline Zwikel, Gabriel Arenas-Henriquez, Wen-jie Ma for discussions. The work of MMShJ is in part supported by the INSF grant No 4026712. The work of HA is supported
by Beijing Natural Science Foundation under Grant No. IS23018 and by the National Natural Science Foundation of China under Grant No. 12150410311.
\appendix

\section{More on different wavefront topologies }\label{sec:topologies}

In this section, we analyze further the symplectic form, charges, and their algebras for different topologies of the wavefront.

\paragraph{Spherical topology.}\label{sec:Spherical-topology}
The sphere case corresponds to $\varepsilon=1$ in metric \eqref{metric-001} and genus $\mathrm{g}=0$. Since we exclude string-like sources and only deal with smooth functions on $S^2$,  fields admit an expansion in terms of spherical harmonics $\mathrm{Y}_{lm}=\mathrm{Y}_{lm}(\theta,\phi)$, 
\begin{subequations}
    \begin{align}
         \lambda^2
    :=&  {\cal R}^2(v) \sum_{l=0}^{\infty} \sum_{m=-l}^{l} \ {\ell}^2_{l,m}(v)\ \mathrm{Y}_{lm}\, , \\
    \Omega  := &\sum_{l=0}^{\infty} \sum_{m=-l}^{l} \ \Omega_{l,m}(v)\ \mathrm{Y}_{lm}\, ,\\  
      \mathrm{P} := &\sum_{l=0}^{\infty} \sum_{m=-l}^{l} \   {\mathrm{P}}_{l,m}(v)\ \mathrm{Y}_{lm}\, .
    \end{align}
\end{subequations}
Then \eqref{R2-def} immediately fixes ${\ell}^2_{0,0}=1/\sqrt{4\pi}$. 
The symplectic form can be written as
\begin{equation}
    \bO_{\text{\tiny reg.}}[g; \delta g]= \delta \beta_\circ \wedge \delta \mathcal{M}_\circ  +\frac{1}{16\pi G} \sum_{l=0}^{\infty} \sum_{m=-l}^{l} \ \delta \Omega_{l,m} \wedge \delta {\mathrm{P}}_{l,m}\, .
\end{equation}
Inverting the above, we can read brackets among modes
\begin{equation}
    \left\{ \beta_\circ, \mathcal{M}_\circ\right\}=1 \, , \qquad \left\{ \Omega_{l,m}, {\mathrm{P}}_{l',m'}\right\}=16 \pi G\, \delta_{l,l'}\delta_{m,m'}\, .
\end{equation}

\paragraph{Toroidal topology.}\label{sec:Toroidal-topology}
This is the simplest case and one can set $z=(\vartheta+ \tau \varphi)/2$ and exploit the identity \eqref{Torus-identity}. To this end, we consider the Fourier series expansion of fields
\begin{subequations}
    \begin{align}
        \lambda^2 :=&\ \mathcal{R}^2 \sum_{m,n \in \mathbb{Z}} \ell^2_{m,n}\, e^{i(m\vartheta+n\varphi)}\, ,\\
        \Omega :=&\  \sum_{m,n \in \mathbb{Z}} \Omega_{m,n}\, e^{i(m\vartheta+n\varphi)}\, ,\\
        \mathrm{P} :=&\  \sum_{m,n \in \mathbb{Z}} \mathrm{P}_{m,n}\, e^{-i(m\vartheta+n\varphi)}\, ,
    \end{align}
\end{subequations}
with $\ell^2_{0,0}=[\pi \mathrm{Im}(\tau)]^{-1}$.
The symplectic form for this case reads
\begin{equation}
    \bO_{\text{\tiny reg.}}[g; \delta g]= \delta \beta_\circ \wedge \delta \mathcal{M}_\circ  +\frac{\pi}{4 G} \sum_{m,n \in \mathbb{Z}} \delta \Omega_{m,n} \wedge \delta {\mathrm{P}}_{m,n}\, ,
\end{equation}
and hence the commutation relations are
\begin{equation}
    \left\{ \beta_\circ, \mathcal{M}_\circ\right\}=1 \, , \qquad \left\{ \Omega_{m,n}, {\mathrm{P}}_{m',n'}\right\}=\frac{4G}{\pi}\, \delta_{m,m'}\delta_{n,n'}\, .
\end{equation}

\paragraph{Higher genus topology.}\label{sec:Higher-genus-topology}
For this case, we can use hyperbolic harmonics which satisfy 
\begin{equation}
    \partial_\psi(\sinh\psi\partial_\psi \Phi_\mathrm{g})+\frac1{\sinh^2\psi}\partial^2_\phi \Phi_\mathrm{g}=-\lambda_\mathrm{g}\ \Phi_\mathrm{g}\, ,
\end{equation}
with $\Phi_\mathrm{g}(\psi,\phi)=\Phi_\mathrm{g}(\psi, \phi+2\pi)$ and $\psi\in [0,\cosh^{-1}(2\mathrm{g}-1)]$. 
One may verify that $\Phi_\mathrm{g}(\psi,\phi)= \Phi_{\lambda_\mathrm{g}, m}(\psi) e^{im\phi}$, where
\begin{equation}
    -\partial_\psi(\sinh\psi\partial_\psi \Phi_{\lambda_\mathrm{g}, m})+\frac{m^2}{\sinh^2\psi}\Phi_{\lambda_\mathrm{g}, m}=\lambda_\mathrm{g}\ \Phi_{\lambda_\mathrm{g}, m}\, ,\qquad m\in \mathbb{Z}\, ,
\end{equation}
are set of orthonormal complete basis on the genus $\mathrm{g}$ surface.

In terms of the above basis, scalar functions may be expanded as
\begin{subequations}\label{z-expand}
\begin{align} 
    \Omega  := &\sum_{{\lambda_\mathrm{g}, m}} \ \Omega_{\lambda_\mathrm{g}, m}(v)\  \Phi_{\lambda_\mathrm{g}, m}(\psi)e^{im\phi}\, ,\\ 
    \lambda^2
    := & {\cal R}^2(v) \sum_{{\lambda_\mathrm{g}, m}} \ {\ell}^2_{\lambda_\mathrm{g},m}(v)\ \Phi_{\lambda_\mathrm{g}, m}(\psi) e^{im\phi}\, , \\ 
    {\mathrm{P}} := &\sum_{{\lambda_\mathrm{g}, m}} \ {\mathrm{P}}_{\lambda_\mathrm{g},m}(v)\ \Phi_{\lambda_\mathrm{g}, m}(\psi) e^{im\phi}\, ,
\end{align}
\end{subequations}
with $\ell^2_{0,0}=1/\sqrt{4\pi}$.
The symplectic form for this case reads
\begin{equation}
    \bO_{\text{\tiny reg.}}[g; \delta g]= \delta \beta_\circ \wedge \delta \mathcal{M}_\circ  +\frac{4G}{\pi } \sum_{{\lambda_\mathrm{g}, m}} \delta \Omega_{{\lambda_\mathrm{g}, m}} \wedge \delta {\mathrm{P}}_{{\lambda_\mathrm{g}, m}}\, .
\end{equation}
and hence the commutation relations are
\begin{equation}
    \left\{ \beta_\circ, \mathcal{M}_\circ\right\}=1 \, , \qquad \left\{ \Omega_{{\lambda_\mathrm{g}, m}}, {\mathrm{P}}_{{\lambda'_\mathrm{g}, m}}\right\}=\frac{\pi}{4G}\, \delta_{m,m'}\delta_{\lambda_\mathrm{g},\lambda'_\mathrm{g}}\, .
\end{equation}

In the above, we did not consider the modular property of the function. To include that for the $\mathrm{g}\geq 1$ cases we could have used Jacobi or Riemann Theta-functions for the expansion \cite{G-R-handbook, Whittaker-Watson}.

		\bibliographystyle{fullsort.bst}
		\bibliography{reference}
	\end{document}